%% file: MPIPublication.tex
\documentclass[12pt]{iopart}
\usepackage{graphicx}
\usepackage{subfigure}
\usepackage{mathptmx}
\usepackage{lineno}
\usepackage{xspace}
\usepackage{upgreek}
\usepackage{mathrsfs}
\usepackage{textcomp}
\usepackage{booktabs}
\usepackage{multirow}
\usepackage{atlasphysics}
\usepackage{preprintcover}  
\usepackage{bm}
\usepackage{cite}
\usepackage{preprintcover}  % load package
\PreprintCoverPaperTitle{Measurement of hard double-parton interactions in $\bm{W (\rightarrow l \nu)}$~+ 2 jet events at $\bm{\sqrt{s}=}$~7~TeV with the ATLAS detector}  % paper title
\PreprintIdNumber{CERN-PH-EP-2012-355}  % CERN preprint number
\PreprintCoverAbstract{The production of $W$ bosons in association with two jets in proton--proton collisions at a centre-of-mass energy of $\sqrt{s}\,=\,7~\TeV$ has been analysed for the presence of double-parton interactions using data corresponding to an integrated luminosity of 36~\ipb, collected with the ATLAS detector at the LHC.  The fraction of events arising from double-parton interactions, $\fDP{D}$, has been measured through the \pt~balance between the two jets and amounts to $\fDP{D} = 0.08~\pm~0.01~(\rm{stat.})~\pm~0.02~(\rm{sys.})$ for jets with transverse momentum $p_{\rm T} > 20$~GeV and rapidity $|y|<2.8$. This corresponds to a measurement of the effective area parameter for hard double-parton interactions of $\sigma_\mathrm{eff} = 15 \pm 3~(\rm{stat.})~^{+5}_{-3} ~(\rm{sys.})~\textrm{mb}$. }  % paper abstract
\PreprintJournalName{NPJ}  % journal name
\def\ptmax{\ensuremath{{p}_{\mathrm{T}}^{\mathrm{max}}}}
\newcommand{\done}{\mathrm{d}}
\newcommand{\sigeff}{\sigma_\mathrm{eff}}
\newcommand{\sigdpi}{\hat\sigma_{Y+Z}^{\mathrm{(DPI)}}}

\newcommand{\sigtot}{\hat\sigma_{Y+Z}^{\mathrm{(tot)}}}
\newcommand{\sigYZ}{\hat\sigma_{Y+Z}^{\mathrm{(SPI)}}}
\newcommand{\sigY}{\hat\sigma_{Y}}
\newcommand{\sigZ}{\hat\sigma_{Z}}

\newcommand{\Eqref}[1]{Equation~(\ref{#1})}

\newcommand{\Secref}[1]{Section~\ref{#1}}

\newcommand{\Figref}[1]{Figure~\ref{#1}}

\newcommand{\Tabref}[1]{Table~\ref{#1}}

\newcommand{\dblp}{double-parton }
\newcommand{\mulp}{multi-parton }

\newcommand{\xsec}{cross section }
\newcommand{\xsecs}{cross sections }
\newcommand{\pileup}{pile-up\xspace}
\newcommand{\Pileup}{Pile-up\xspace}
\newcommand{\com}{centre-of-mass\xspace}
\newcommand{\DeltaJ}{\ensuremath{\Delta_{\mathrm{jets}}}}
\newcommand{\DeltaJn}{\ensuremath{\Delta_{\mathrm{jets}}^\mathrm{n}}}
\newcommand{\fDP}[1]{\ensuremath{f_{\mathrm{DP}}^\mathrm{(#1)}}}

\newcommand{\AHJ}{\textsc{A+H+J}\xspace}

\begin{document}
%\linenumbers
%\preprint{Submitted to NPJ} 

\title[Hard double-parton interactions in $W (\rightarrow \ell  \nu)$~+ 2-jet events]{Measurement of hard double-parton interactions in $\bm{W (\rightarrow \ell  \nu)}$~+ 2-jet events at $\bm{\sqrt{s}=}$~7~TeV with the ATLAS detector}
\author{The ATLAS Collaboration} 
%\email[Full author list given at the end of the article. ]{}

\begin{abstract}
The production of $W$ bosons in association with two jets in proton--proton collisions at a centre-of-mass energy of $\sqrt{s}\,=\,7~\TeV$ has been analysed for the presence of double-parton interactions using data corresponding to an integrated luminosity of 36~\ipb, collected with the ATLAS detector at the LHC.  The fraction of events arising from double-parton interactions, $\fDP{D}$, has been measured through the \pt~balance between the two jets and amounts to $\fDP{D} = 0.08~\pm~0.01~(\rm{stat.})~\pm~0.02~(\rm{sys.})$ for jets with transverse momentum $p_{\rm T} > 20$~GeV and rapidity $|y|<2.8$. This corresponds to a measurement of the effective area parameter for hard double-parton interactions of $\sigma_\mathrm{eff} = 15 \pm 3~(\rm{stat.})~^{+5}_{-3} ~(\rm{sys.})~\textrm{mb}$. 
\end{abstract}

\maketitle
\section{Introduction} 
\label{sec:intro}
\input{intro.tex}
\section{Theoretical background}
\label{sec:theory}
\input{theory.tex}
\section{Strategy of the analysis} 
\label{sec:strategy}
\input{strategy.tex}

\section{The ATLAS detector}  
\label{sec:det}
\input{detector.tex}
\section{Event selection}
\label{sec:select}
\input{event_selection.tex}

\section{MC simulation}
\label{sec:mc}
\input{mc.tex}
\section{Strategy validation}
\label{sec:MCVal}
\input{validation.tex}

\section*{Extraction of $\bm{\fDP{D}}$}
\label{sec:measurement}
\input{measurement.tex}

\section{Hadron-level studies}
\label{sec:hadlevel}
\input{hadronlevel.tex}

\section*{Determination of $\bm{\sigeff}$}
\label{sec:sigeff}
\input{sigeff.tex}
\section{Conclusions}
\input{conclusions.tex}

\input{Acknowledgements.tex}
\clearpage

\include{bib}
\onecolumn
\clearpage
\input{atlas_authlist} 
\end{document}

%% file: intro.tex
Double-parton interactions (DPI) in hadron-initiated processes have been 
discussed in theoretical studies since the first days of the parton 
model~\cite{Landshoff:1978fq,Takagi:1979wn,Goebel:1979mi}.  These studies have 
subsequently been refined and reformulated in the framework of perturbative 
quantum chromodynamics (QCD) for a variety of processes such as double Drell--Yan production, four-jet production, and $W$ production associated with two 
jets~\cite{Paver:1982yp,Mekhfi:1983az,Humpert:1983fy,Humpert:1984ay,Ametller:1985tp,Halzen:1986ue,Godbole:1989ti}. Potential correlations in colour and spin space have been 
analysed theoretically~\cite{Mekhfi:1985dv}, and evolution equations for \mulp
distribution functions have been derived~\cite{Shelest:1982dg}.
The formalism ~\cite{Humpert:1984ay,Ametller:1985tp} to deal with double-parton interactions in hadronic interactions at a centre-of-mass energy $\sqrt{s}$ may be summarised, assuming perturbative factorisation, by

\begin{equation}
\label{Eq::mpi}
\eqalign{\done\sigdpi(s) = &&\frac{m}{2\sigma_{\rm{eff}}(s)} \int \done x_{i_1}\; \done x_{j_1}\; \done x_{i_2}\;\done x_{j_2}\; [f_{i_1j_1}(x_{i1},\,x_{j_1},\,\mu_F) \cr
&&f_{i_2j_2}(x_{i_2},\,x_{j_2},\,\mu_F)\;\done\hat\sigma_{i_1i_2\to Y}(x_{i_1},\,x_{i_2},\,s)\;\done\hat\sigma_{j_1j_2\to Z}(x_{j_1},\,x_{j_2},\,s)],}
\end{equation} 

\noindent where $\done\sigdpi$ is the differential \dblp interaction \xsec for the inclusive production of 
a combined system $Y+Z$ at a given $\sqrt{s}$, and the $\done \hat\sigma_{k_1 k_2\to Y(Z)}$ 
is the differential partonic cross section for the production of a system $Y$ or $Z$ in the 
collision of partons $k_1$ and $k_2$. The symmetry factor $m$ is equal to one if $Y=Z$ and equal to two if $Y\neq Z$. The $f_{ij}(x_{i},\,x_{j},\,\mu_F)$ are the 
double-parton distribution functions (DPDFs) evaluated at a specific 
factorisation scale, $\mu_F$. The integration over the momentum fractions $x_{i}$ and $x_{j}$ of the two partons from the same proton is constrained by 
energy conservation such that $x_{i}+x_{j}\le 1$. A summation over all possible parton combinations is implicitly assumed. 
%Folded with parton distribution functions 
%(PDFs), they would give rise to hadronic cross sections $\hat\sigma_Y$.  
%Note that in any case, the cross sections are completely differential in the
%$n$-body final state phase space of system $Y$, later denoted by $\Phi_Y$.  
Typically, the DPDFs are expressed in terms 
of the conventional single parton distributions using a factorised 
\textit{ansatz}~\cite{Humpert:1984ay,Ametller:1985tp}, namely

\begin{equation}
\label{Eq::DPDF_factorised}
f_{ij}(x_{i},\,x_{j},\,\mu_F) = f_i(x_{i},\,\mu_F)\,f_j(x_{j},\,\mu_F)\;(1-x_{i}-x_{j})\; \Theta\; (1-x_{i}-x_{j})\,,
\end{equation}

\noindent where the factor $(1-x_{i}-x_{j}) \Theta(1-x_{i}-x_{j})$ implements the kinematic constraint and $\Theta(x)$ is the Heavyside step function. The effective area parameter for \dblp interactions, $\sigma_{\rm{eff}}(s)$, is defined at the parton level and, in the 
formalism outlined here, is independent of the process and of the phase-space under consideration.
Naively, it can be related to the geometrical size of the proton, leading to an estimate of $\sigeff\approx \pi R_p^2 \approx 50$~mb, where $R_p$ is the proton radius.  
Alternatively, $\sigeff$ can be connected to the inelastic cross section,
which would lead to $\sigeff\approx\sigma_{\rm{inel}}\approx 70$~mb at 
$\sqrt{s}=7~\TeV$~\cite{Tompkins:2011bv,Antchev:2011vs}.  

A number of measurements of $\sigeff(s)$ have been performed in $pp$ or $p\bar p$ collisions at \com energies of 63~GeV~\cite{Akesson:1986iv}, 630~GeV~\cite{Alitti:1991rd},
1.8~TeV~\cite{Abe:1993rv,Abe:1997xk} and
1.96~TeV~\cite{Abazov:2009gc}. The measured values range from about  5~mb at the lowest energy to about 15~mb at Tevatron energies. Attempts to understand these values have used non-trivial correlations between the two scattering systems to explain the 
differences between these measured values~\cite{Calucci:1997uw,Calucci:1999yz}. 

In the scientific programme of the Large Hadron Collider (LHC), issues related to
\mulp interactions have attracted increasing attention ~\cite{Ryskin:2011kk,Kom:2011nu,Gaunt:2009re,Diehl:2011tt,Drees:1996rw,DelFabbro:1999tf, Kulesza:1999zh,DelFabbro:2002pw,Hussein:2006xr,Maina:2009vx,Maina:2009sj,Berger:2009cm,Gaunt:2010pi,Maina:2010vh}. 
%This includes general considerations concerning the form~\cite{Ryskin:2011kk} and 
 %     calculation~\cite{Gaunt:2009re} of the DPDFs, the structure of the emerging analytical expressions~\cite{Gaunt:2011xd}, correlations in \mulp interactions~\cite{Diehl:2011tt} and other phenomenological studies~\cite{Drees:1996rw,DelFabbro:1999tf,
%      Kulesza:1999zh,DelFabbro:2002pw,Hussein:2006xr,Maina:2009vx,Maina:2009sj,
  %    Berger:2009cm,Gaunt:2010pi,Maina:2010vh,Kom:2011nu}.
This surge of interest is due to the higher \com energy leading to enhanced parton densities
and therefore to an anticipated larger impact of such effects on a multitude
of physics signatures. 
The high energy and high luminosity available at the LHC also implies that multiple
interactions should occur at higher transverse momentum, $\pt$, offering
the possibility to further study these interactions in a variety of
processes.
 %Assuming a simple, Regge-physics inspired power 
%law of the type $\sigeff(s)\propto s^\kappa$ yields a scaling behaviour of 
%this parameter with $\kappa\approx 0.12$; a more careful analysis was 
%presented in \cite{Domdey:2009bg}. \\
This paper presents a measurement of $\sigeff$ in $pp$ collisions at $\sqrt{s}=7~\TeV$ performed with the ATLAS detector~\cite{ATLAS}, using events with two jets produced in association with a $W$ boson.
%Although $\sigeff$ appears to be non-perturbative in nature, through its connection to %non-%perturbative 
%quantities such as the proton radius or inelastic cross sections, the 
%formalism above applies to systems $Y$ and $Z$ which are 
%produced at scales that are hard enough to allow for a perturbative treatment.
%This is reflected in the choice of final states in the measurements quoted
%above, and it is also true in this paper, where the parameter $\sigeff$ is determined 
%using the transverse momentum, $\pt$, imbalance of the two jets.

%% file: theory.tex
The quantity $\sigeff$ parameterises 
the \dblp interaction part of the production \xsec for a composite system 
$(Y+Z)$ in hadronic collisions.  Assuming no correlations between the two systems, the differential \xsec $\sigtot$ 
for the production of $Y+Z$ consists of a direct part, $\sigYZ$, originating from single-parton interaction, and a \dblp 
interaction contribution, $\sigdpi$,

\begin{equation}
\label{Eq::master}
%\begin{split}
\done\sigtot(s) \;=\; \done\sigYZ(s)+\done\sigdpi(s) 
\;=\; \done\sigYZ(s)+\frac{\done\sigY(s)\cdot\done\sigZ(s)}{\sigeff(s)}\,,
%\end{split}
\end{equation}

\noindent where $\done\sigY(s)$ and $\done\sigZ(s)$ correspond to the differential cross sections of processes $Y$ and $Z$ respectively and the symmetry factor $m$ from \Eqref{Eq::mpi} has been set equal to two.

After integrating the differential cross sections in \Eqref{Eq::master} over the phase space defined by the selection cuts on the $Y$ and $Z$ systems as appropriate for the analysis and solving the equation for $\sigma_{\rm{eff}}$, 

\begin{equation}
\label{equation_def_sigeff}
\sigeff(s) \;=\; 
\frac{\sigY(s)\cdot\sigZ(s)}{\sigdpi(s)}\;=\;
\frac{\sigY(s)\cdot\sigZ(s)}{\sigtot(s)-\sigYZ(s)}\,.
\end{equation}

\noindent With the exception of the direct component, $\sigYZ(s)$, for which theoretical input (for example
in the form of a Monte Carlo -- MC -- event generator) needs to be employed, all quantities in \Eqref{equation_def_sigeff} may be directly taken from data, provided that the simple factorisation picture is applicable. 
This assumes that the proposed correlation in the 
DPDFs, the factor $(1-x_{i}-x_{j})$ present in \Eqref{Eq::DPDF_factorised}, 
is close to 1. There 
will be other effects which will eventually lead to a breakdown of this 
simple picture in some corners of phase space; for example total energy 
conservation, flavour conservation rules, or, more intricately, complicated 
interactions between the initial- or final-state partons \cite{Blok1,Blok2,Gaunt1,Gaunt:2009re,Gaunt3} which potentially correlate the two systems in a
non-trivial way.  However, for certain processes and selection cuts, such 
effects may turn out to be negligible. In the following, when referring to data and integrating over the hadronic final states, $\hat\sigma(s)$ will be replaced by $\sigma(s)$.  

\begin{figure}[t]
\begin{center}
\mbox{
\includegraphics[height=0.20\textheight]{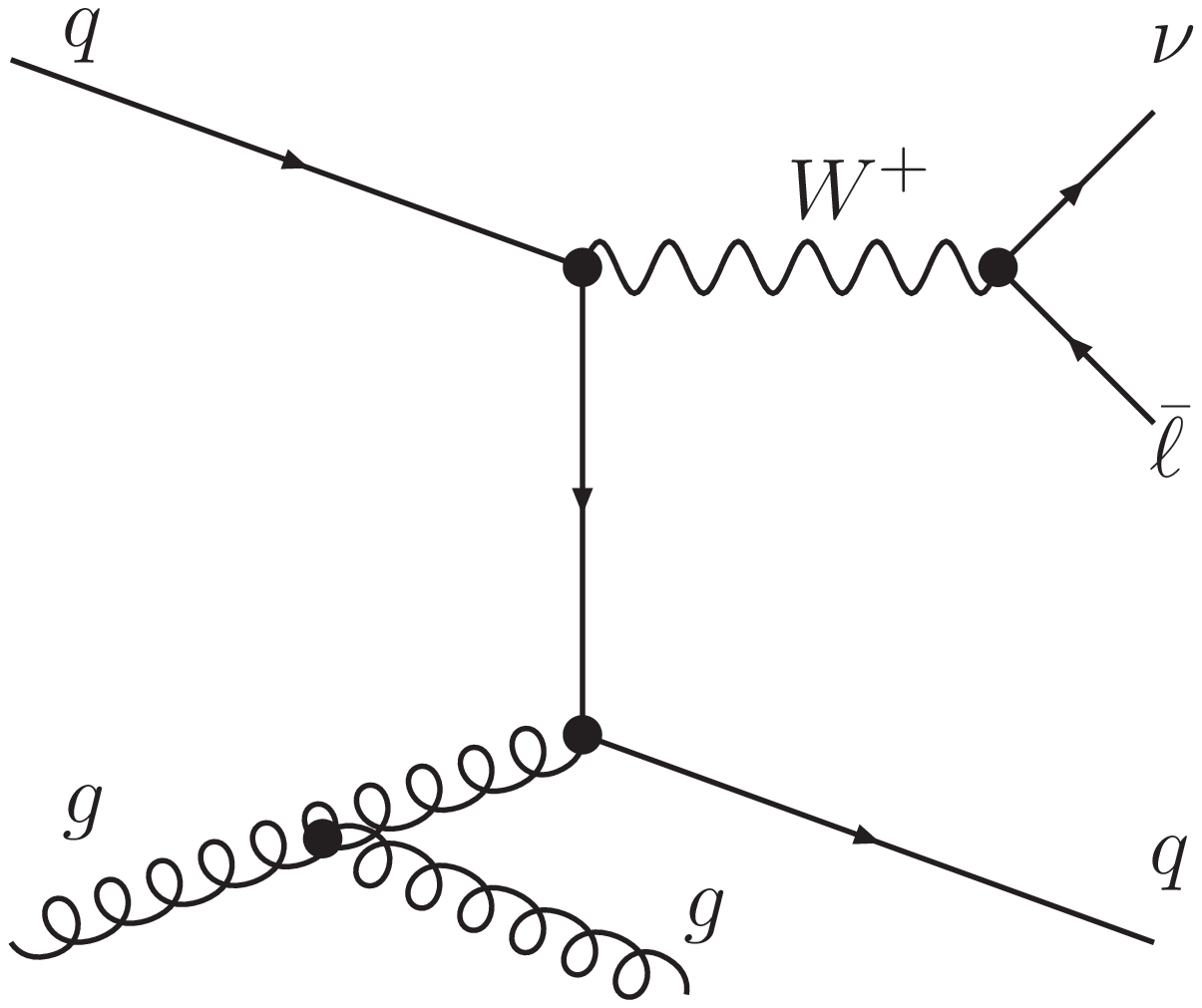} 
\includegraphics[height=0.20\textheight]{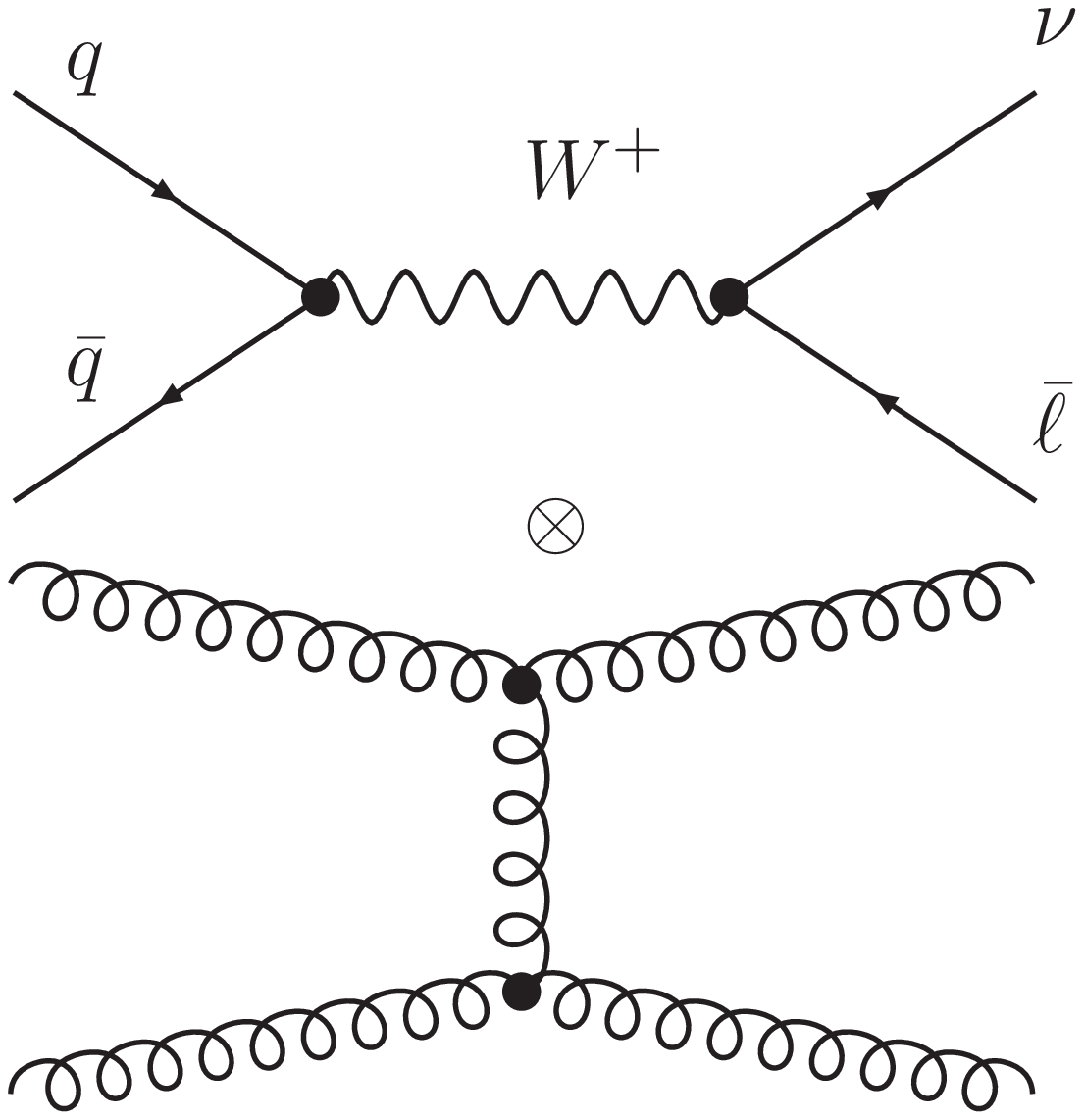} 
}
\caption{Examples of leading-order Feynman diagrams for the direct (left) and 
  \dblp interaction (right) components in the production of a $W^++$2-jet 
  system. These contributions are defined in \Eqref{Eq::master} with 
  the identification $Y\to W^+$ and $Z\to 2$ jets.}
\label{Fig::Feyns}
\end{center}
\end{figure}

For the case of $W+$2-jet production discussed in this paper, the cross sections at leading order are related to 
Feynman diagrams such as those depicted in \Figref{Fig::Feyns}.  In
general, calculations of the differential cross sections for the 
production of any system $Y$ are inclusive. In 
particular, in the calculation of the leading-order $W$ cross 
section, the production of additional jets is implicitly included. These extra jets may populate a phase space constrained from below by
the factorisation scale $\mu_{\rm{F}}$. In this study, in which jets are defined by a transverse momentum requirement of $\pt~>20$~GeV, this implies that the cross sections entering the calculation of $\sigeff$ correspond to the production of a $W$ boson with no accompanying jets, of a $W$ boson accompanied by exactly two jets and a hadronic final state consisting of exclusive two jets.

%% file: strategy.tex
\label{Analysis_strategy} 
\label{sec:fitmethod} 
\label{sec:observable} The cornerstone of the analysis is the extraction of the fraction of $W+$2-jet events produced in $pp$ interactions in which the jets originate from a hard DPI. This fraction is subsequently used to determine the value of $\sigeff$. The sample of $W+$2-jet candidate events is selected from data recorded by the ATLAS detector with the $W$ boson identified through its leptonic decays into $e\nu$ and $\mu\nu$. The fraction $\fDP{D}$, where the superscript $\mathrm{D}$ refers to detector level, of $W+$2-jet events originating from DPI is defined by

\begin{equation}
\label{equation_defC}
\fDP{D} \;=\; \frac{N_{W_{0\mathrm{j}}+2\mathrm{j_{DPI}}}}{N_{W+2\mathrm{j}}}\;=\;
\frac{N_{W_{0\mathrm{j}}+2\mathrm{j_{DPI}}}}{N_{W_{2\mathrm{j}}}\;+\;N_{W_{0\mathrm{j}}+2\mathrm{j_{DPI}}}},
\end{equation}
\noindent where $N_{W+2\mathrm{j}}$ is the total number of $W+$2-jet events, $N_{W_{2\mathrm{j}}}$ is the number of events in which the production of the two jets is directly associated with the production of the $W$ boson (single-parton interaction), and $N_{W_{0\mathrm{j}}+2\mathrm{j_{DPI}}}$ is the number of events in which the production of the two jets originates from DPI. 
In order to extract $\fDP{D}$, a minimisation fit to the distribution of an observable 
is performed. The observable is chosen such that it shows good discriminating power between the direct production of a $W$ boson with two jets ($W_{2\mathrm{j}}$) and the production of a $W$ boson in association with zero jets in addition to another parton--parton scatter resulting in two jets ($W_{0\mathrm{j}}+2\mathrm{j_{DPI}}$). The fit is performed to the normalised, detector-level, background-corrected data distribution of the observable using two normalised templates, denoted by $A$ and $B$. \\
\indent Template $A$ represents the expected contribution to the distribution of the chosen observable from $W_{2\rm{j}}$ events, and Template $B$ that from $W_{0\mathrm{j}}+2\rm{j_{DPI}}$ events. The fit function is given by
$(1-\fDP{D})\cdot A+\fDP{D} \cdot B$.
%, where the fit 
%parameter $\fDP{D}$ corresponds to the coefficient multiplier of the 
%DPI template and thus the fraction of $W$+DPI contamination in the data.  
The details of how the templates are constructed and how the fit is performed, 
together with supporting MC studies, are described in
\Secref{sec:mc}.
The relevant part of the equation defining $\sigeff$,
\Eqref{equation_def_sigeff}, reads

\begin{equation}\centering
\label{equation_Sigeff}
\sigeff\;=\;\frac{\sigma_{W_{0\mathrm{j}}}\cdot \sigma_{2\rm{j}}}{\sigma_{W_{0\mathrm{j}}+2\rm{j_{DPI}}}}\,.
\end{equation}

\noindent Here, $\sigma_{W_{0\mathrm{j}}}$, $\sigma_{W_{0\mathrm{j}}+2\mathrm{j_{DPI}}}$ and $\sigma_{2\mathrm{j}}$ are the production
\xsecs of $W_{0\rm{j}}$, $W_{0\rm{j}}+2\rm{j_{DPI}}$ and exclusive dijet (2j) 
events, respectively.  These \xsecs are related to the 
respective number of events $N$ through the relation

\begin{equation}
\sigma \;=\; \frac{N}{{\cal{A}}~C~\epsilon~\mathscr{L}}\,,
\end{equation}

\noindent where $C$ denotes the corrections for unfolding to the particle level including reconstruction effects, 
$\cal{A}$ is the geometrical acceptance, $\epsilon$ is the trigger efficiency, and 
$\mathscr{L}$ is the integrated luminosity.  
%Thus, \Eqref{equation_Sigeff}, becomes
%\begin{equation}
%\label{equation_intermedSigeff} 
%\sigeff\;=\;
%\frac{1}{\fDP{D}}\cdot\frac{N_{W_0}~N_{2\rm{j}}}{N_{W+2\rm{j}}} \cdot 
%\frac{C_{W_0+2\rm{j_{DPI}}}}{C_{W_0}~C_{2\rm{j}}} \cdot 
%\frac{A_{W_0+2\rm{j_{DPI}}}}{A_{W_0}~A_{2\rm{j}}} \cdot 
%\frac{\epsilon_{W_0+2\rm{j_{DPI}}}}{\epsilon_{W_0}~\epsilon_{2\rm{j}}} \cdot 
%\frac{\mathscr{L}_{W_0+2\rm{j_{DPI}}}}{\mathscr{L}_{W_0}~\mathscr{L}_{2\rm{j}}}.
%\end{equation}
The assumption of factorisation between the $W$ boson and the $2\rm{j}$ system leads to some simplifications.  
%This leads to a number of implications regarding the quantities in 
%Eq.~\eqref{equation_intermedSigeff}:
First, the kinematics of the $W$ boson does not influence the kinematic 
distributions of the DPI system, either at the detector level or at the hadron 
level, once corrections involving the impact of jets on $W$ reconstruction and vice 
versa have been made.
% as described in section \ref{sec:sigeff}.  
Secondly, the kinematics of the jets originating from DPI may be modelled by the 
kinematics of single-scatter dijet events. Therefore

\begin{equation}
\label{equation_effCancellation}
{\cal{A}}_{W_{0\mathrm{j}}+2\rm{j_{DPI}}} \cdot C_{W_{0\mathrm{j}}+2\rm{j_{DPI}}}\;=\;
{\cal{A}}_{W_{0\mathrm{j}}} \cdot C_{W_{0\mathrm{j}}} \cdot {\cal{A}}_{2\rm{j}} \cdot C_{2\rm{j}}.
\end{equation}

Finally, the $W_{0\mathrm{j}}+2\rm{j_\mathrm{DPI}}$ and $W_{0\mathrm{j}}$ events are collected
using the same trigger selection. Taken together, this results in luminosity and efficiency cancellations and 
$\sigeff$ is given by
%Eq.~\eqref{equation_intermedSigeff} simplifies, quite dramatically, to 

\begin{equation}
\label{equation_sigeff} 
\sigeff\;=\;
\frac{N_{W_{0\mathrm{j}}}~N_{2\rm{j}}} {\fDP{D}\cdot{N_{W+2\rm{j}}}}
\cdot 
\frac{1}{\mathscr{\epsilon}_{2\rm{j}}} \cdot \frac{1}{\mathscr{L}_{2\rm{j}}}.
\end{equation}

In a previous phenomenological study~\cite{Godbole:1989ti} it was suggested to use the transverse momentum of the $W$ boson, $\pt^W$, as the key 
observable to distinguish \dblp scattering production of $W+$2-jet events 
from the direct production channel.  This observable 
suffers from experimental inaccuracies due to the fact that the kinematics of 
the $W$ boson must be reconstructed from the missing transverse momentum, \met.  Alternatively, one could try 
to use the $\pt$ distribution of the individual jets, but their discrimination power is
limited by uncertainties stemming from the jet energy scale.  This leaves
correlations between the jets or between the jets and the kinematics of the $W$ as
further possibilities, one being the azimuthal correlation of the two
leading jets in the transverse plane.  In the picture of DPI 
production advocated here, the kinematics of the $W$ boson and the dijet systems 
are decorrelated. Therefore, the momenta of the
two jets must compensate each other in the transverse plane, orienting them
back-to-back in azimuthal angle, rendering, in principle, this angular separation between the jets a useful 
observable. However, due to the distortion of this variable by various systematic effects, in particular multiple proton--proton interactions (\pileup) and the underlying event, in this analysis the balance in transverse momenta of the two jets is used instead, quantified by 

\begin{equation}
  \label{Eq:DefDelta}
  \DeltaJ\;=\;|{\vec{p}_\mathrm{T}}^{J_1}+{\vec{p}_\mathrm{T}}^{J_2}|\,,
\end{equation}

\noindent where ${\vec{p}_\mathrm{T}}^{J_1}$ and ${\vec{p}_\mathrm{T}}^{J_2}$ are the transverse momentum vectors of the two leading jets.
Anticipating a potentially large dependence of  $\DeltaJ$ on the jet energy scale, another 
observable, the transverse momentum of the dijet system normalised by the
sum of the individual transverse momenta, $\DeltaJn$, is constructed,

\begin{equation}
  \label{Eq:DefDelta_n}
  \DeltaJn\;=\;
  \frac{|{\vec{p}_\mathrm{T}}^{J_1}+{\vec{p}_\mathrm{T}}^{J_2}|}{|\vec{p}_\mathrm{T}^{J_1}|+|\vec{p}_\mathrm{T}^{J_2}|}\,.
\end{equation}

The distribution of $\DeltaJn$ is employed to drive the fit from which 
$\fDP{D}$ is obtained, while the distribution of $\DeltaJ$ allows further 
checks.

%% file: detector.tex
The ATLAS detector~\cite{ATLAS} comprises a superconducting solenoid surrounding the inner detector (ID), as well as electromagnetic and hadronic calorimeters and a large superconducting toroid magnet system instrumented with muon-detection chambers. The ID system is immersed in a 2~T axial 
magnetic field and provides tracking information for charged particles in a 
pseudorapidity range matched by the precise measurements of the 
electromagnetic calorimeter. The silicon pixel and microstrip tracking 
detectors cover the pseudorapidity\footnote{ATLAS uses a right-handed 
          coordinate system with its origin at the nominal interaction point 
          (IP) in the centre of the detector and the $z$-axis along the beam 
          pipe. The $x$-axis points from the IP to the centre of the LHC ring, 
          and the $y$-axis points upward.  Cylindrical coordinates $(r,\phi)$ 
          are used in the transverse plane, $\phi$ being the azimuthal angle 
          around the beam pipe. The pseudorapidity is defined in terms of the 
          polar angle $\theta$ as $\eta=-\ln\tan(\theta/2)$. The rapidity of 
          a particle with respect to the beam axis is defined as 
          $y = \frac{1}{2}\ln\frac{E+p_z}{E-p_z}$.}
range $|\eta| < 2.5$. The transition radiation tracker, which surrounds 
the silicon detectors, can perform tracking up to $|\eta| = 2.0$ and contributes 
to electron identification.
The liquid-argon electromagnetic calorimeter is divided into one 
barrel ($|\eta| < 1.475$) and two end-cap components ($1.375 < |\eta| < 3.2$). 
It uses an accordion geometry to ensure fast and uniform response, and fine 
segmentation for optimum reconstruction and identification of electrons and 
photons.  The iron/scintillator tile hadronic calorimeter consists of a barrel 
covering the region $|\eta| < 1.0$, and two extended barrels in the range 
$0.8 < |\eta| < 1.7$. 
The muon spectrometer is based on three large superconducting toroids with 
coils arranged in an eight-fold symmetry around the calorimeters, covering a 
range of $|\eta|< 2.7$.  Over most of this range, precision measurements of 
the track coordinates in the principal bending direction of the magnetic field 
are provided by monitored drift tubes.  At large pseudorapidities 
($2.0 < |\eta| < 2.7$), cathode strip chambers with higher granularity 
are used in the innermost station.  
The ATLAS detector has a three-level trigger system consisting of level-1,
level-2 and the event filter (L1, L2 and EF).  The L1 trigger rate at design 
luminosity is approximately 75 kHz.  The L2 and EF triggers reduced the recorded event rate in 2010  to approximately 200 Hz. 

%% file: event_selection.tex
\label{sec:evsel}
The dataset collected in 2010, corresponding to approximately 36\;\ipb~of integrated 
luminosity, is used in this analysis. The rate of overlapping uncorrelated proton--proton interactions occurring within the same bunch crossing (also referred to as pile-up) gradually increased up an average of about two interactions per bunch-crossing throughout the data-taking period. 
The selection of $W$ events is based on the leptonic $\Wen$ and $\Wmn$ decay channels and follows the one already described in Ref.~\cite{WjetsXS}, where more details can be found. It differs in the requirements on the jet transverse momentum and rapidity. The objects required for the different samples and templates needed in the analysis were selected as described in the following.

To select electron candidates, clusters formed~from energy depositions in the electromagnetic calorimeter are required with matched tracks in the inner detector, with the~further requirement that the cluster shapes are consistent with electromagnetic showers initiated by electrons. Such electron candidates were required to have transverse momenta $\pt^\ell\,>\,20$~GeV and $|\eta|<2.47$. Electrons reconstructed in the transition region between the barrel and end-cap calorimeters ($1.37<|\eta|<1.52$) or falling into inactive regions 
  of the calorimeter were excluded.  The standard isolation requirement on electron 
  candidates~\cite{egamma} was applied to improve multi-jet background rejection.

Muon candidates were selected by requiring $\pt^\ell\,>\,20$~GeV and $|\eta|<2.4$.  
  They were reconstructed requiring both the muon spectrometer and 
  the inner detector information.  Additional requirements were applied to the number of 
  hits used to reconstruct the track in the inner detector.  Furthermore, the 
  $z$-coordinate of the muon longitudinal impact parameter with respect to the interaction 
  vertex was required to be less than 10~mm.  A selection requirement was applied to the 
  significance of the track transverse impact parameter to ensure that the muon was 
  prompt. The standard isolation requirements were applied to the muons~\cite{muons} to improve multi-jet background rejection.

 To select events with a $W$ boson, in addition to requiring exactly one lepton ($e$ or $\mu$) in the
  event, requirements were imposed on $\met$ and the transverse mass, $m_{\rm T}$.  The $\met$ was calculated using the reconstructed physics objects, the remaining energy deposits in the 
  calorimeter, and the inner detector tracking information.  The $m_{\rm T}$ is defined in terms of the missing 
  transverse energy and the charged lepton transverse momentum as 
  $m_{\rm T} = [2\pt^\ell \met(1-\cos\Delta\phi_{\ell,E_{\rm T}^{\rm miss}})]$, where 
  $\Delta\phi_{\ell,\met}$ is the angle between the lepton transverse momentum and
  the missing transverse momentum.  Events were required to have $\met > 25$~GeV and $m_{\rm T} > 40$~GeV.

 Jets are defined using the anti-$k_t$ algorithm~\cite{antikt} 
  with radius parameter $R=0.4$. The jets were reconstructed 
  from clusters built from calorimeter cells, initially not accounting for different calorimeter response to electrons and hadrons, and subsequently calibrated~\cite{LCCali1}.  Jets were required 
  to have $\pt>20$~GeV and $|y|<2.8$.  All jets within $\Delta R = \sqrt{(\Delta\eta)^2+(\Delta\phi)^2} = 0.5$ of a 
  selected electron or muon were removed from the analysis.
  The number of jets originating from \pileup 
  interactions was reduced by applying a selection requirement on the jet-vertex fraction (JVF), 
  which is defined for each jet in an event.  After associating tracks to 
  jets with a matching in $\Delta R$(track, jet), requiring $\Delta R < 0.5$, 
  the JVF was computed for each jet as the scalar sum of the transverse momenta
  of all matched tracks from the interaction vertex divided by the total 
  jet-\-matched track transverse momentum from all vertices.  Jets were removed 
  from the analysis if $\mathrm{JVF} < 0.75$.  This selection was not applied to 
  jets that lie outside the acceptance of the inner detector or to those jets 
  without matching tracks.

With these selections, the following samples have been constructed:
\begin{itemize}
\item{$W+$0-jet sample, consisting of events passing either the $\Wen$ or $\Wmn$ selection and where no jets are found in addition to the $W$ decay products; 
}
\item{$W+$2-jet sample, consisting of events passing either the $\Wen$ or $\Wmn$ selection and where exactly two jets are found in addition to the $W$ decay products;
} 
\item{dijet sample, which consists of events recorded with exactly two jets. The selected events were taken using the Minimum Bias Trigger Scintillators and Zero Degree Calorimeters, which have been shown~\cite{jetpaper} to be unbiased and fully efficient for jet-based measurements. A subset of this sample with negligible \pileup, corresponding to the first $184~\upmu \rm{b}^{-1}$ of data taking, was used to calculate the $N_{\rm{2j}}$ term described in \Eqref{equation_sigeff}.}
\end{itemize}

%% file: mc.tex
The treatment of MC simulation and background estimation is based on that described in Ref.~\cite{WjetsXS}. The MC samples of events in this analysis were produced with CTEQ6L1~\cite{CTEQ} parton distribution functions. The simulation of detector effects was performed with \textsc{Geant4}~\cite{geant}.  These simulated event samples \cite{atlassim} were used for the background and signal estimates.  They were reweighted such that they matched the data in the number of reconstructed vertices per event. 

To simulate the $W$ boson signal, samples of events were generated using \textsc{Alpgen}~\cite{alpgen} with the MLM~\cite{mlm} matching scheme, interfaced with \textsc{Herwig}~\cite{herwig} v6.510 and \textsc{Jimmy}~\cite{jimmy} v4.31 (AUET2 tune), together referred to as \AHJ samples. \textsc{Sherpa}~\cite{Sherpa} v1.3.1, with the CKKW~\cite{ckkw} merging, and with the default underlying event tune, was also used to simulate $W$ events.

%To arrive at a DPI-off sample in \textsc{Alpgen+Herwig+Jimmy} (\AHJ), events 
%with additional hard partonic interactions were removed from the inclusive sample.  This removal was %achieved by analysing the structure of the MC event record.  To arrive at a "DPI-off" sample (defined in %Section \ref{section_templates}), events containing two or more additional $2\to 2$ parton scatters %with a parton-level $\pt\ge\ptmax = 15\;$GeV, defined before the jet finder was run, were removed %(see also Section \ref{section_templates}).  

MC generators populate the activity in a hard-scattering event with additional parton--parton scatters, the average number of which depends on the assumed lowest-$\pt$ threshold of additional soft partons and the available phase space. The extra scatters are commonly called multi-parton interactions (MPI). Some of the scattered partons will materialise as jets of hadrons above a certain $\pt$ threshold. Others will result in extra hadronic activity added to the original hard scattering in the event. 
In this analysis, a threshold on the $\pt$ of the scattered parton, $\ptmax$, is introduced, whose purpose is to separate these two classes of hard and soft scatters.

In the \AHJ simulation, a sample of events without DPI may be obtained by removing events with two or more additional $2\to 2$ parton scatters with a parton-level $\pt$ above a preset $\ptmax$.  In \textsc{Sherpa}, the \textsc{MI\_HANDLER} switch is used to switch off the effects of DPI. 

%In this case, all secondary parton scatterings above a scale of 3.5\;GeV were removed. This implies %that the ``DPI off'' sample of \textsc{Sherpa} contains physics different from the one in the %corresponding \AHJ ``DPI-off'' sample. 
As for potential sources of physics background to the $W$ signal, \textsc{Pythia6}~\cite{pythia} was used to produce MC samples for the modelling of multi-jet and $Z\rightarrow\ell\ell$ physics contributions.
\textsc{Powheg}~\cite{powheg} was used to model the $t\overline{t}$ 
background contribution. MC@NLO~\cite{mcanlo} was used to provide MC samples 
for the modelling of single top quark and diboson contributions. 

\subsection{Sample composition} 
\label{section_bg}
%Backgrounds to the selected events come from other types of electroweak events (Z → ll and W → τν), tt ̄ and single top events, and from multijet events in which a non- prompt lepton is either produced through the a coincidence of hadronic signatures that mimics the characteristics of a lepton.

 The contributions of electroweak backgrounds ($Z\rightarrow \ell \ell$ and diboson production), as well as $W\rightarrow\tau\nu$, $t\overline{t}$ and single top quark production to both channels are estimated using Monte Carlo simulation. The absolute normalisation is derived using the total theoretical cross sections and corrected using the acceptance and efficiency losses of the event selection. These backgrounds amount to about 5\% of the selected events in the electron channel (dominated by the $W\rightarrow\tau\nu$ contribution), and about 8\% in the muon channel (dominated by the $Z\rightarrow\ell\ell$ contribution). The contributions from non-physics backgrounds were considered to be negligible.
The background contributions to the selected events can also come from multi-jet production processes in which a lepton is either produced through the decay of a hadron containing a heavy quark, the decay-in-flight of a light meson to a muon, or through a coincidence of hadronic signatures mimicking the characteristics of a lepton. The shape and normalisation of the distribution of various observables in multi-jet backgrounds are determined using data-driven methods in both analysis channels. For the $\Wen$ selection, the background shape is obtained by reversing certain requirements on shower shape in the calorimeter in the data selection procedure to produce a multi-jet enriched sample. Similarly, to estimate the multi-jet contribution to $\Wmn$, the background shape is obtained from data by inverting the requirements on the muon impact parameter and its significance to produce a multi-jet enriched sample.
These multi-jet--enriched samples give the shapes of the distributions of multi-jet background observables. Their normalisation in the selected data sample is determined by fitting a linear combination of the multi-jet \met~ shape, and that for the leptonic contribution, to the observed \met~ distribution. The multi-jet background was thus estimated to contribute about 14\% of the selected events in the electron channel and 6\% in the muon channel in the selected $W$+2-jet sample.

% In both cases, the relative normalisation of the template is determined by fitting the \met~ distribution in the data (without the \met~ selection requirement) to a sum of two templates: one for the multijet background and another which includes the signal and the leptonic backgrounds. An additional correction for the acceptance, derived from simulation, is also applied. 

\begin{figure*}[t]
\centering
\mbox{
\subfigure[]{\includegraphics[width=0.5\textwidth]{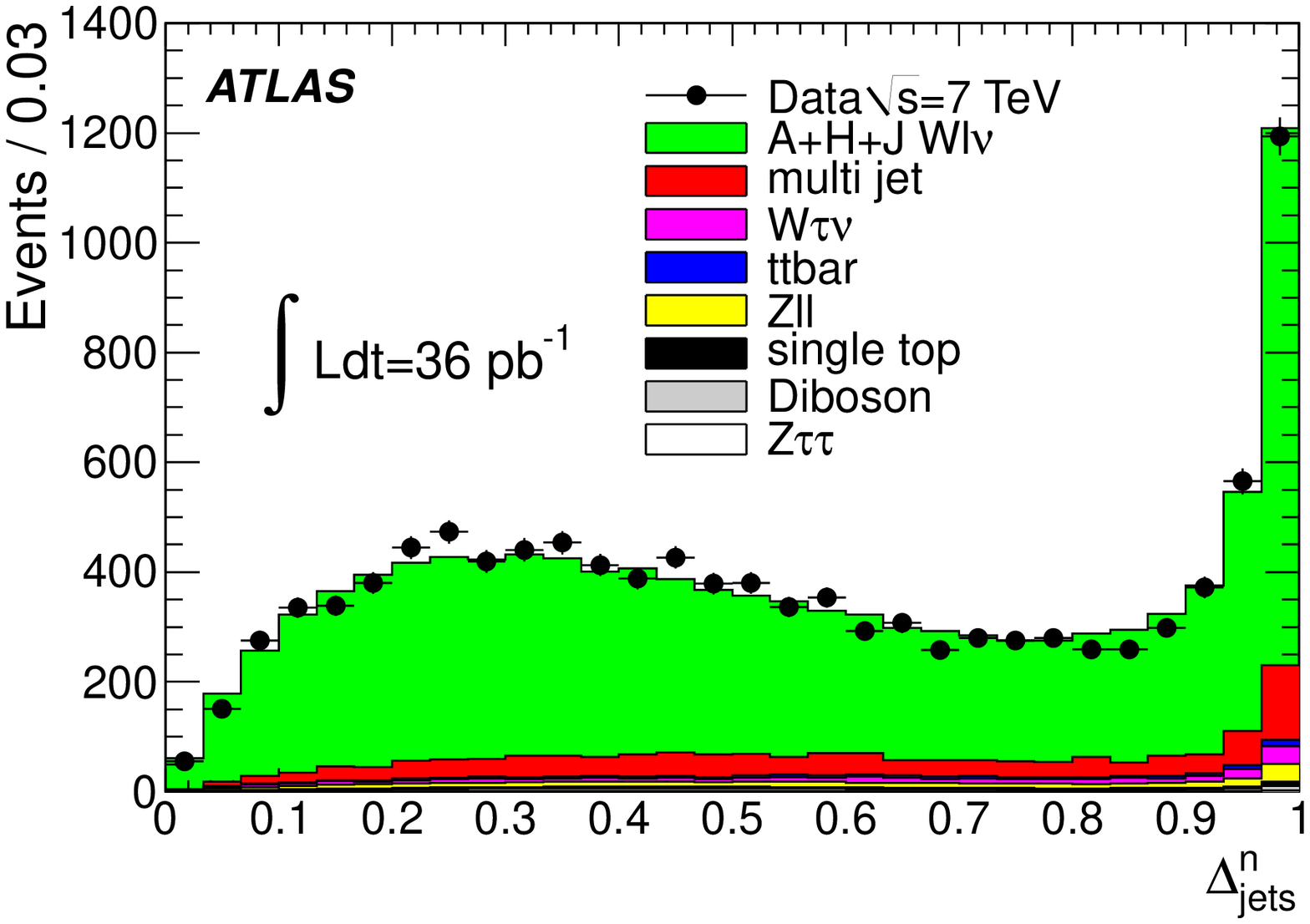}}
\subfigure[]{\includegraphics[width=0.5\textwidth]{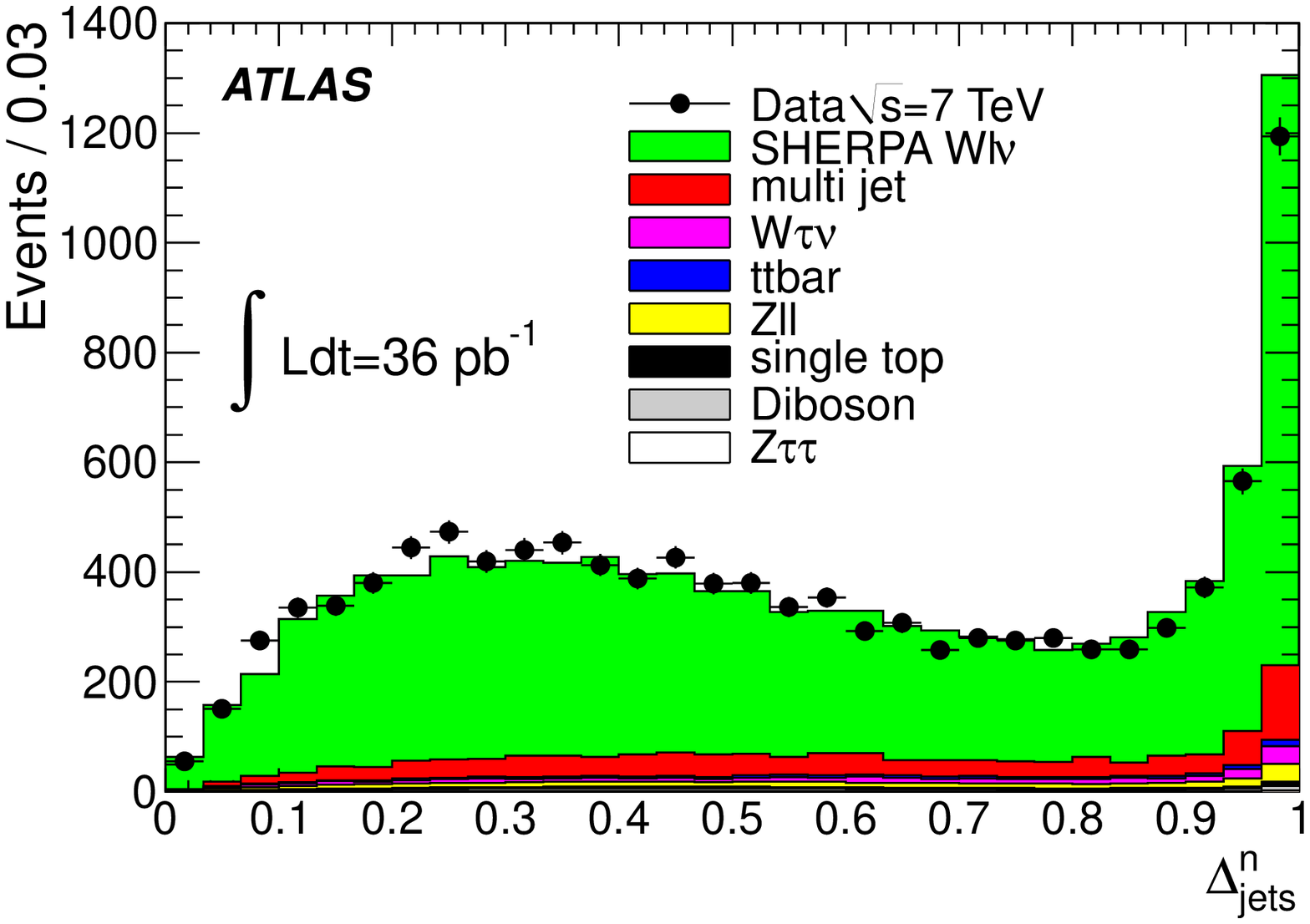}}
}
\mbox{
\subfigure[]{\includegraphics[width=0.5\textwidth]{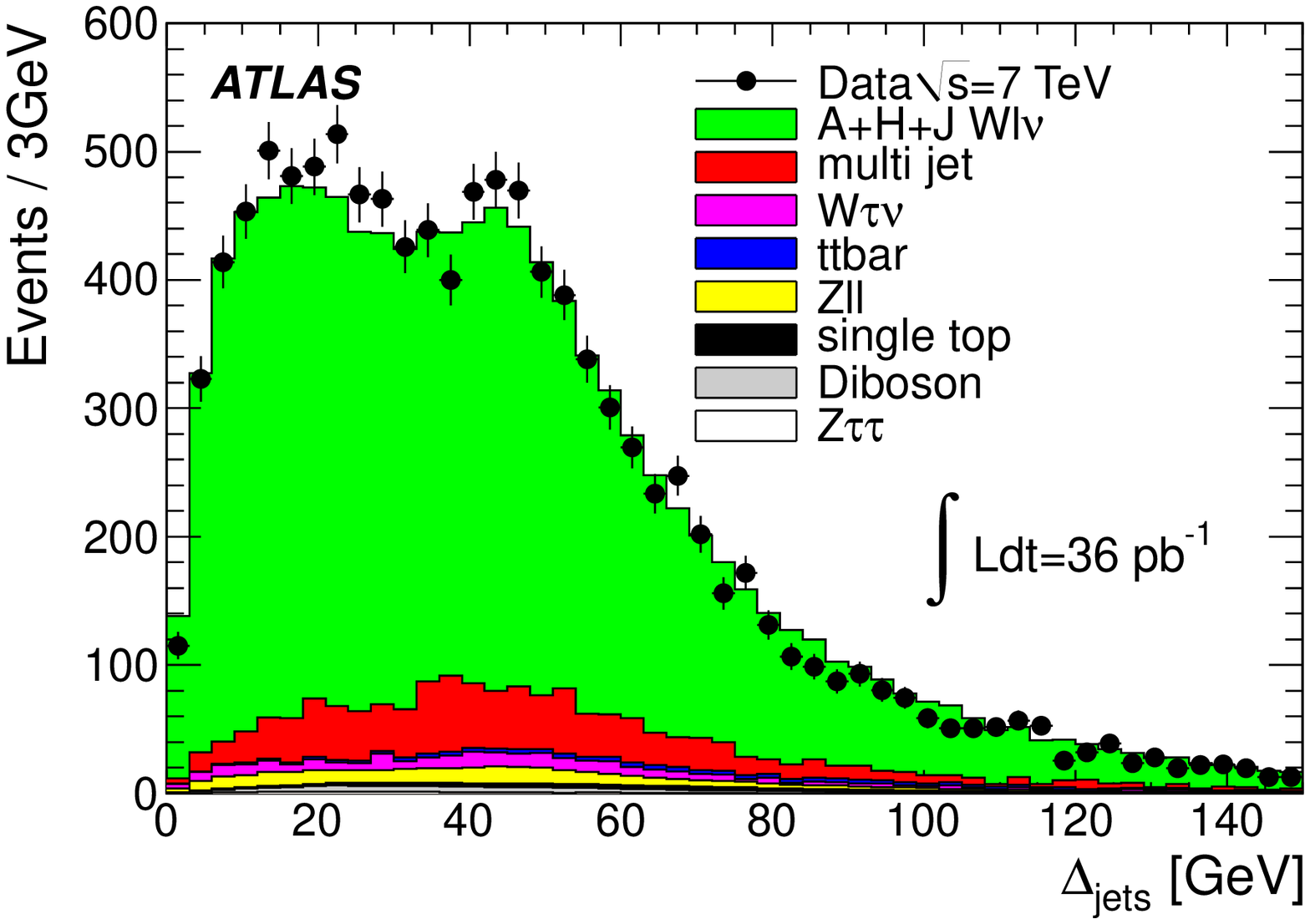}}
\subfigure[]{\includegraphics[width=0.5\textwidth]{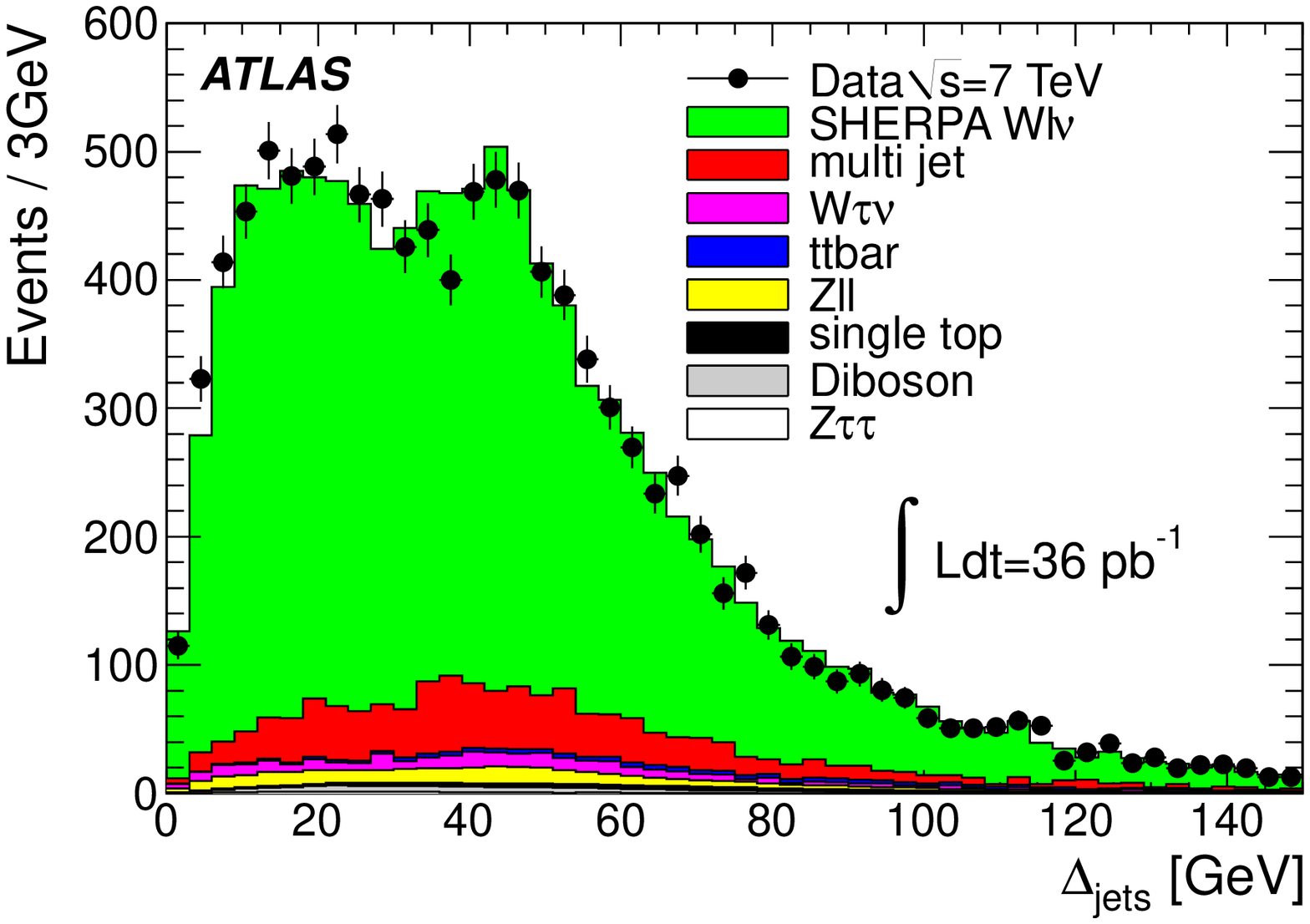}}
}
\caption{The (a, b) $\DeltaJn$ and (c, d) $\DeltaJ$ 
  distributions at detector level for events passing the $W$+2-jet 
  selection cuts.  The distributions from data (dots) are compared with 
 (a, c) \AHJ and (b, d) \textsc{Sherpa} signal 
  MC (histogram) predictions.  In addition, physics backgrounds, also shown, have
  been added in due proportion to the MC histogram.}
\label{figure_sigBG}
\end{figure*}

\Figref{figure_sigBG} shows the distributions of the two key observables $\DeltaJn$ and
$\DeltaJ$ obtained in selected $W$+2-jet events. The data are compared with 
the results from the \textsc{Sherpa} and \AHJ MC samples with their default MPI treatment, after adding the background contributions, which are also shown in the plots.

\subsection{Templates}
\label{section_templates}
The fit, from which $\fDP{D}$ is extracted, is performed by comparing the distribution of $\DeltaJn$
at detector level in background-corrected data with two templates.
\begin{itemize}
\item Template $A$ ``DPI-off'' --
normalised distribution of the discriminating variable for a sample in which the two jets originate from a primary scatter. The main sample for Template $A$ was produced by \AHJ. To construct Template $A$, it is necessary to remove hard MPI candidate events from the generated sample. Relevant events contain the $W$ boson as well as at least two outgoing partons, {\it i.e.},\ partons that stem from the hardest scatter, or that originate from a secondary scatter.  Events containing two or more secondary partons above a defined cutoff scale, $\ptmax\,=\,15\;$GeV, are classified as DPI events and therefore rejected.  The value of $\ptmax$ and 
its impact on the analysis are further discussed in \Secref{sec:partoncutoff}.\\
An alternative modelling of Template $A$ was obtained using the \textsc{Sherpa} 
MC sample where MPI has been switched off altogether.  This removes 
all secondary perturbative parton scatters, which effectively produce 
transverse momenta in the range $\pt>3.5\;$GeV, but it retains the initial-state radiation off the incoming legs of the hard matrix element, the 
generation of intrinsic transverse momentum and the fragmentation of beam 
remnants. 

As an example, the distributions of $\DeltaJn$ and $\DeltaJ$ for simulated $W$+2-jet events
are displayed in \Figref{figure_MPIeffects2}.  These distributions were obtained 
from \AHJ, both inclusively and with DPI switched off (i.e. Template $A$). The distributions show sensitivity to the effect of \dblp scattering, especially at low values of $\DeltaJn$ and $\DeltaJ$.
%There is a marked difference between \AHJ and \textsc{Sherpa}, which can be attributed to the different treatment of soft multiple scatters, as is discussed in \Secref{sec:MCVal}. 

\begin{figure*}[htbp]
\centering
\mbox{
\subfigure[]{\includegraphics[width=0.5\textwidth]{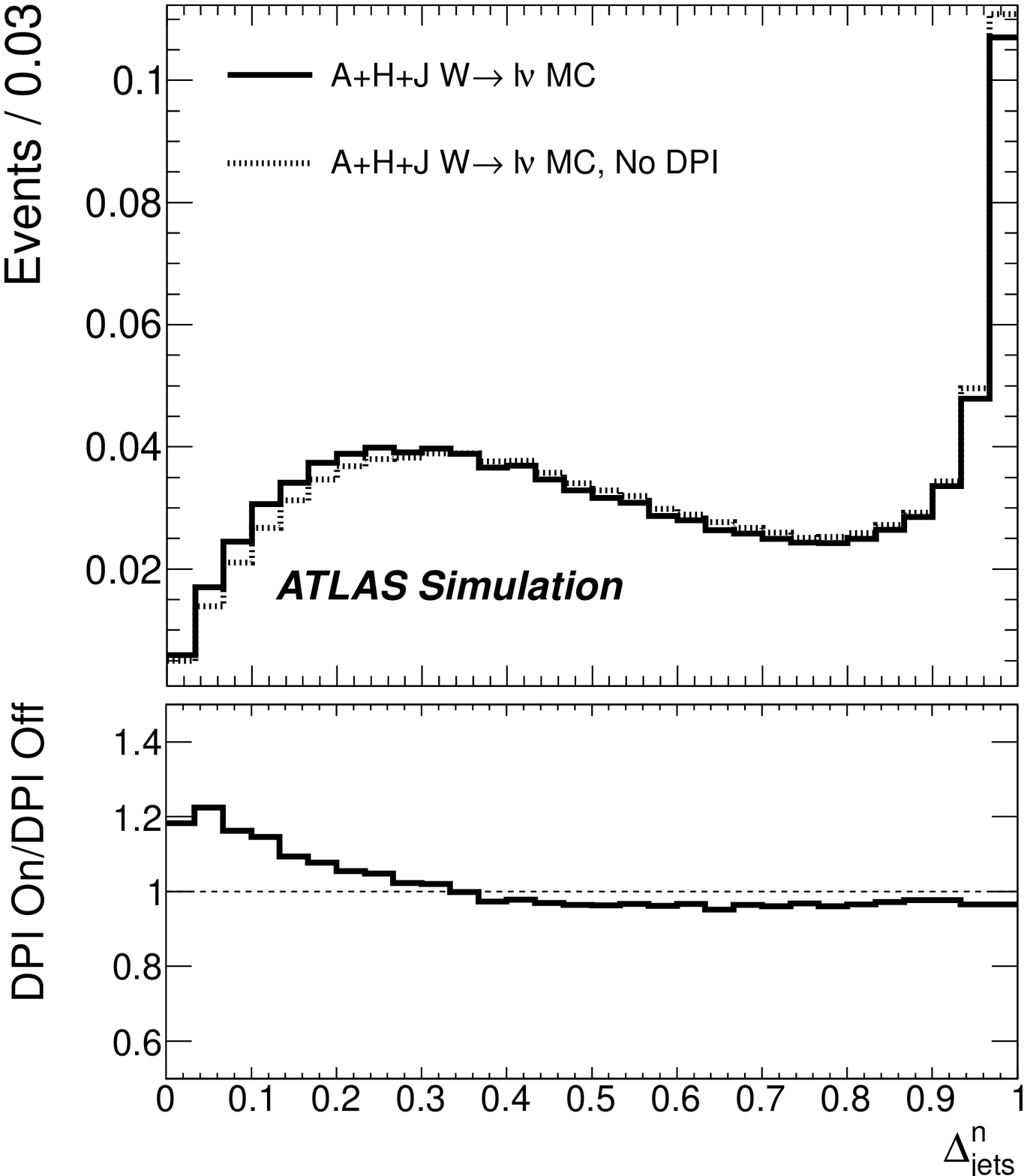}}
\subfigure[]{\includegraphics[width=0.5\textwidth]{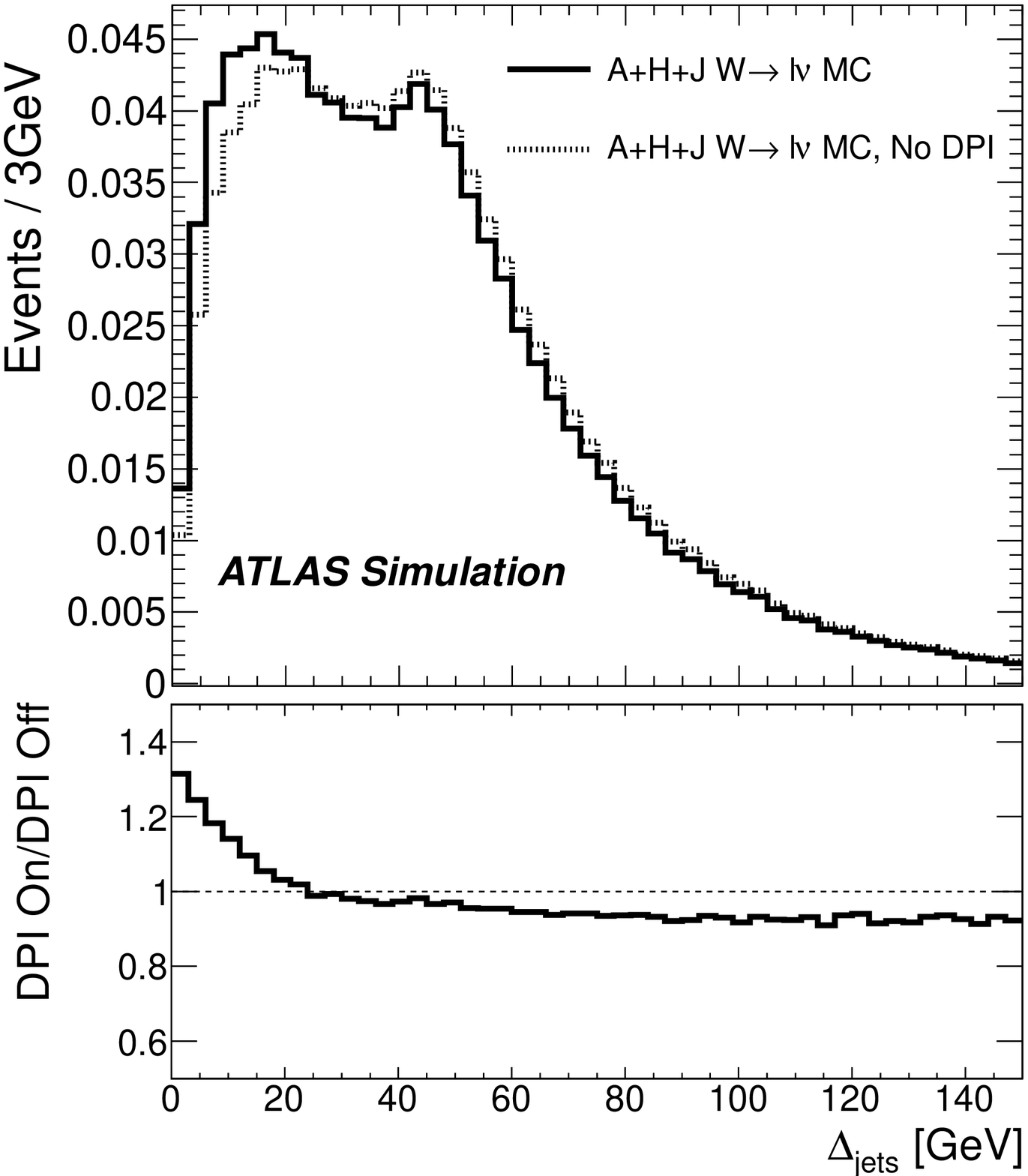}}
}
\caption{Comparison of the shapes of (a) $\DeltaJn$ and (b) $\DeltaJ$ distributions, at 
  detector level, for selected $\Wln+2$-jet events as expected in \AHJ, with DPI on and off (Template $A$). The lower panels show the respective ratios of the DPI-on and DPI-off expectations.}
\label{figure_MPIeffects2}
\end{figure*}

\item Template $B$ ``DPI-only'' -- 
normalised distribution of the discriminating variable for a sample in which both jets originate from a DPI scatter.
The dijet sample described in \Secref{sec:evsel} was used to approximate the DPI template.  The 
fractional difference between the final results when using the dijet 
\textsc{Pythia6} MC as Template $B$ in place of dijet data was found to be well below the percent level.  
\end{itemize}

%% file: validation.tex
For the purpose of validation studies, the inclusive \AHJ sample was employed to mimic the data and its distribution was fitted as a combination of Templates $A$ and $B$. As for the fits to the data distributions, Template $A$ is based on the \AHJ sample with $\ptmax = 15\;$GeV and Template $B$ is based on the dijet data sample. The fitted value for $\fDP{MC}$ thus obtained, once corrected for \pileup effects, can be compared to the expected fraction of DPI directly extracted at the parton level from the event record of the \AHJ sample. 

\subsection{DPI at detector level in the MC simulation}
The fit of the combination of Template $A$ and Template $B$ to the nominal \AHJ $\DeltaJn$ distribution yields
%and represents the DPI rate that would be extracted in the \AHJ MC. 

\begin{equation}
\fDP{MC}(\DeltaJn)\;=\;
  0.051 \pm 0.003\;\;\;\mathrm{(stat.)}.
\end{equation}

\begin{figure*}[t]
\centering
\mbox{
\subfigure[]{\includegraphics[width=0.5\textwidth]{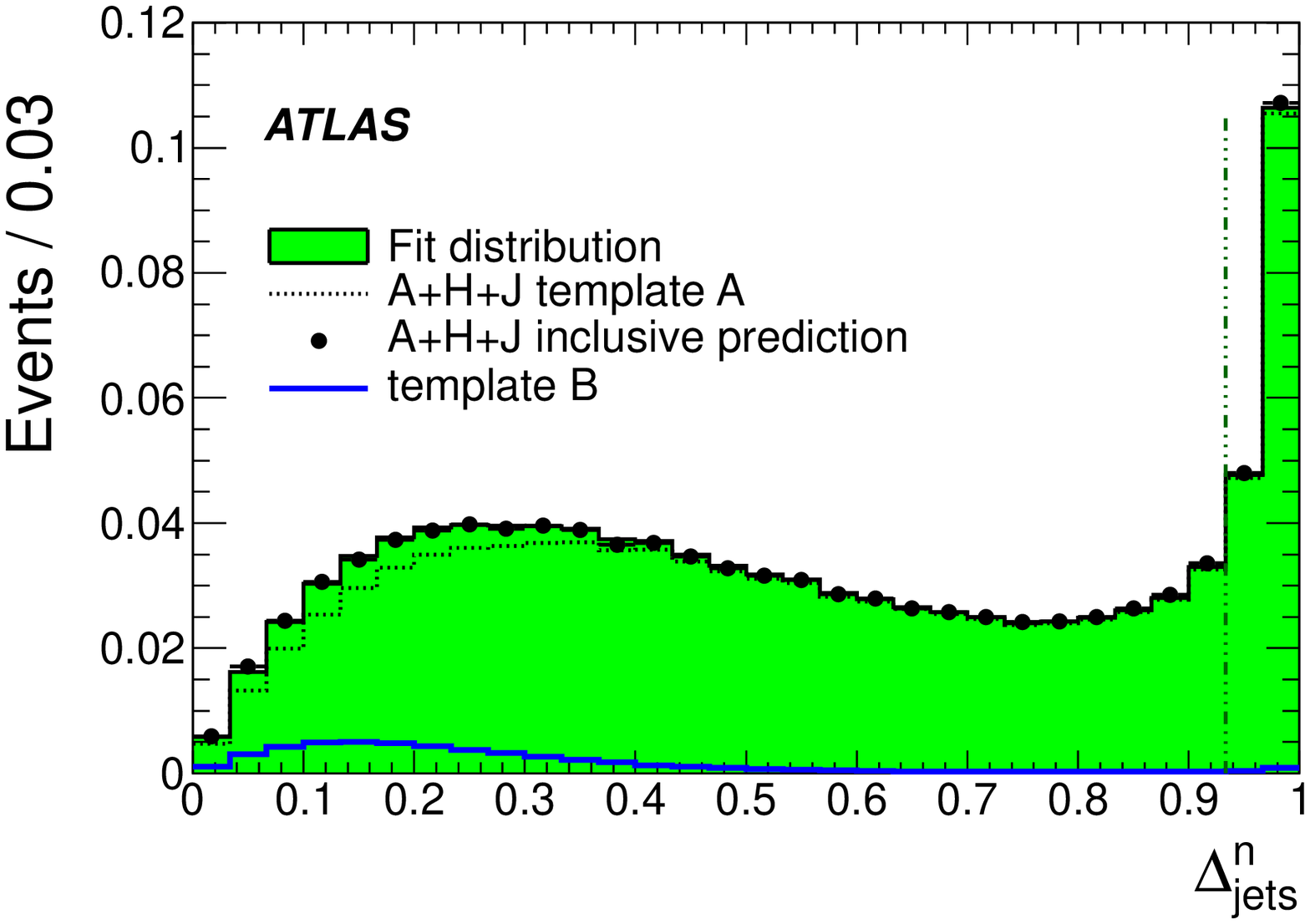}}
\subfigure[]{\includegraphics[width=0.5\textwidth]{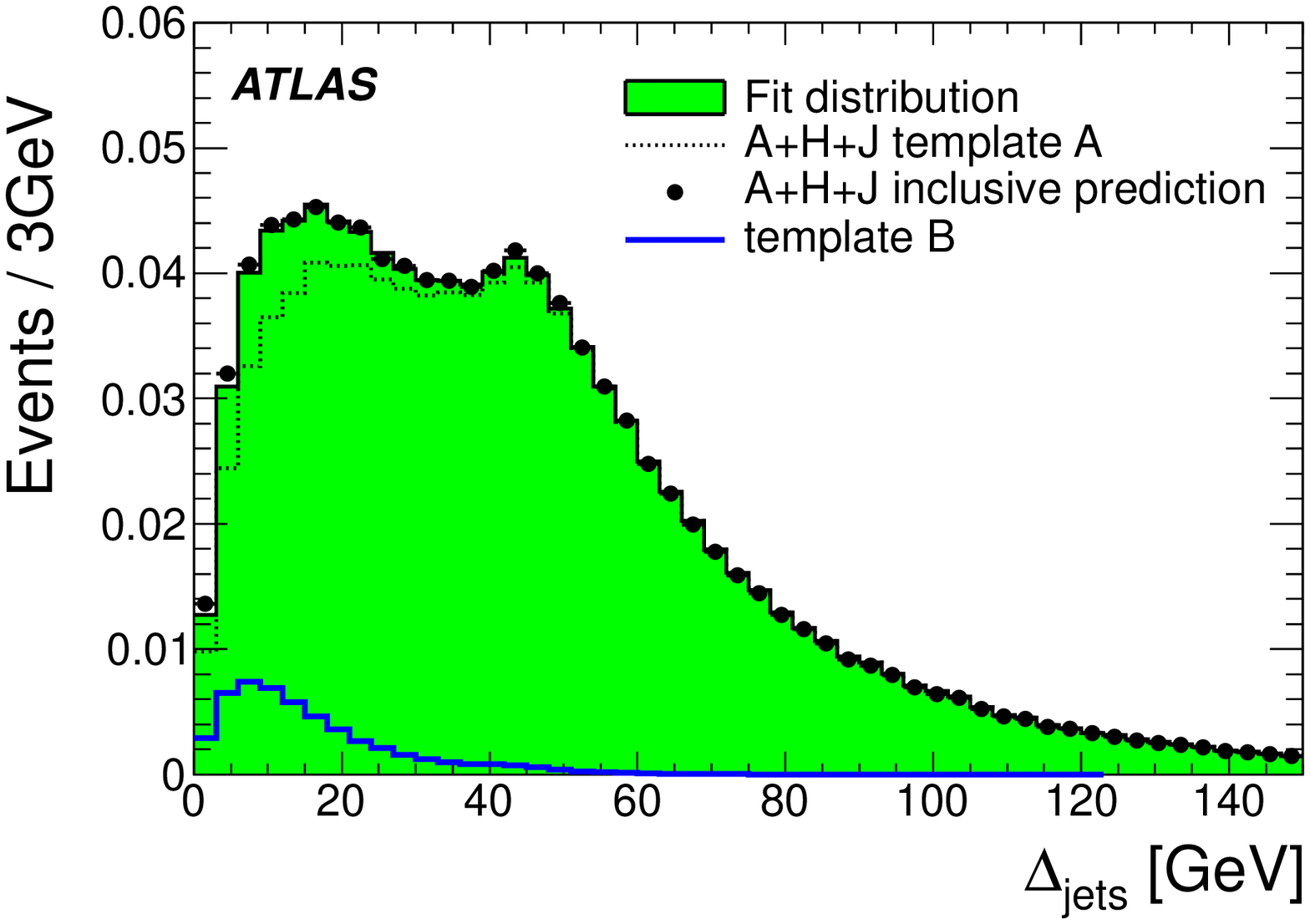}}
}
\caption{Distribution of (a) $\DeltaJn$ and (b) $\DeltaJ$ in the inclusive \AHJ 
  pseudo-data (dots) compared to the results of fitting $\DeltaJn$ by a linear combination 
  of Template $A$ (dashed line) extracted from this sample and of Template $B$ 
  obtained from the dijet data (blue solid line).  The result is shown as
  the green histogram.  The bins to the right of the vertical dash-dotted 
  line were excluded from the fit.  The pseudo-data and the overall fit have been 
  normalised to unity, Template $A$ to $1-\fDP{MC}(\DeltaJn)$ and Template 
  $B$ to $\fDP{MC}(\DeltaJn)$.}
\label{figure_closurefit}
\end{figure*}

In the fit to the distribution of $\DeltaJn$, events with $\DeltaJn\;>\;0.93$ (corresponding to the last two bins of the fit) were ignored, since they represent configurations with two nearly parallel jets and therefore rather test the parton shower model. The fit minimisation, when performed to the $\DeltaJ$ instead of the $\DeltaJn$ distribution, resulted in a value \fDP{MC}(\DeltaJ) that was within 13\% of \fDP{MC}(\DeltaJn).  
The resulting description of the distributions in $\DeltaJn$ and $\DeltaJ$ by the combination of the Templates $A$ and $B$, using 
$\fDP{MC}(\DeltaJn)$, are shown in \Figref{figure_closurefit}.

%To check the validity of the assumption of process and phase space independence of $\sigeff$, its %variation in MC simulation was evaluated under 
%different event selection criteria and, in particular, the \pt~ and $\eta$ 
%requirements on the jets.  While in general largely different values of 
%$\fDP{MC}$ were extracted, the MC equivalent to 
%\Eqref{equation_sigeff} yielded fractional variations in $\sigeff^{\rm{MC}}$ that were smaller than 5\%.

\subsection{Influence of \pileup}
\label{sec:pileupcorr}
In order to account for the influence of \pileup, the extraction of 
$\fDP{MC}$ was repeated after selecting only events with the requirement of 
exactly one reconstructed vertex, imposed on both the inclusive \AHJ 
sample and Template $A$.  A subset of dijet events from earlier data-taking periods, where the effects of \pileup were smaller, was used to model Template $B$.  In this way, 
the fitted value of $\fDP{MC}$ represents the DPI rate that would be extracted in 
the absence of \pileup. The result is

\begin{equation}
  \fDP{MC}(\DeltaJn) \;=\; 
  0.059\pm 0.007\;\;\;\mathrm{(stat.)}\,,
\end{equation}

\noindent which is in good agreement with that obtained from a fit to 
the $\DeltaJ$ distribution. The ratio, $r_\mathrm{\pileup}$, of the $\fDP{MC}$ value with the one-vertex requirement to that without the requirement, without accounting for the effect of correlations\footnote{The impact of including the correlations is estimated to result in a maximum reduction of the statistical uncertainty on $r_\mathrm{\pileup}$ to 0.12.}, is $r_\mathrm{\pileup} = 1.17\pm 0.15\;\mathrm{(stat.)}$. A direct extraction of $\fDP{D}$ using only single $pp$ interactions is not possible due to the small numbers of events in the data. For this reason, the ratio is used to correct $\fDP{D}$ to the result that would have been obtained in the case of single interactions. The statistical uncertainty on $r_\mathrm{\pileup}$ is duly propagated as a systematic uncertainty whenever appropriate.

\subsection{Transition from detector level to parton level}
\label{sec:partonlevel}
An important question, central to this analysis, is whether the $\sigeff$
extracted at detector level can be related to the same quantity at the parton 
level.  This is relevant because the value of $\sigeff$ used in theoretical 
applications  is typically defined at parton level only.
This question can be translated into how close the DPI rate at parton level, 
$\fDP{P}$, is to the extracted DPI rate at detector level in the MC samples, $\fDP{MC}$.  The parton-level DPI fraction is defined as

\begin{equation}
\label{equation_fdp_p}
\fDP{P}\;=\;
\frac{{{N}}^{\rm{P}}(W_{0\rm{j}}+2\rm{j_{DPI}})}
     {{{N}}^{\rm{P}}(W_{0\rm{j}}+2\rm{j_{DPI}})+{{N}}^{\rm{P}}(W_{2\rm{j}})},
\end{equation}
where the various ${N}^{\rm{P}}$ are the number of corresponding events at 
parton level.  Here, the jets are directly identified with outgoing partons, 
with a fiducial acceptance chosen such that it matches that of the
jets at detector level,

\begin{equation}
  \pt^{\rm P}\ge 20\;\textrm{GeV}\;,\;\;\;|y^{\rm P}|\le 2.8
  \;\;\;\mathrm{and}\;\;\;\Delta R_{\mathrm{P}\ell}>0.5\,,
\end{equation}

\noindent where $\ell$ denotes charged leptons and $\mathrm{P}$ the partons, as recorded 
at the generator level.  The parton-level counting of the quantities 
${{N}}^{\rm{P}}(W_{0\rm{j}}+2\rm{j_{DPI}})$ and ${N}^{\rm{P}}(W_{2{\rm{j}}})$ was 
performed in the inclusive \AHJ MC sample after selection of $W$+2-jet events, and yielded
\begin{equation}
\label{fdp_partonlevel}
\fDP{P}(\DeltaJn) = 
0.064\pm0.001\;\;\;\mathrm{(stat.)}, 
\end{equation}

\noindent which is within 10\% of the equivalent quantity at detector level, $\fDP{MC}$. 

\subsection{Effect of \ptmax~value}
\label{sec:partoncutoff}
\noindent There could be sizable 
differences between the DPI-off samples provided by \AHJ and \textsc{Sherpa}. 
The origin of these differences has been identified as the difference in the extent to which softer MPI, at scales below the dijet transverse momenta, are included in the samples.  In \textsc{Sherpa}, the only option 
is to switch off all such secondary interactions, while in \AHJ this can be 
steered through the $\ptmax$ cut. In the \AHJ samples, events are removed from 
the inclusive sample if the $\pt$ in secondary parton--parton scatters is 
above $\ptmax = 15\;$GeV. All softer MPI scatters are left in the sample.  In 
contrast, in the DPI-off \textsc{Sherpa} sample, {\em all} secondary parton 
scatters are switched off, corresponding roughly to a $\ptmax \approx 3.5\;$GeV.  To see how closely 
the two models can agree, $\ptmax$ in \AHJ was reduced to $3.5\;$GeV.  Despite 
intrinsic differences in the MPI modelling, this choice replicates the 
\textsc{Sherpa} results to within 10\%, as is discussed in \Secref{sec:measurement}.

\begin{figure}[t]
\centering
\includegraphics[width=0.8\textwidth]{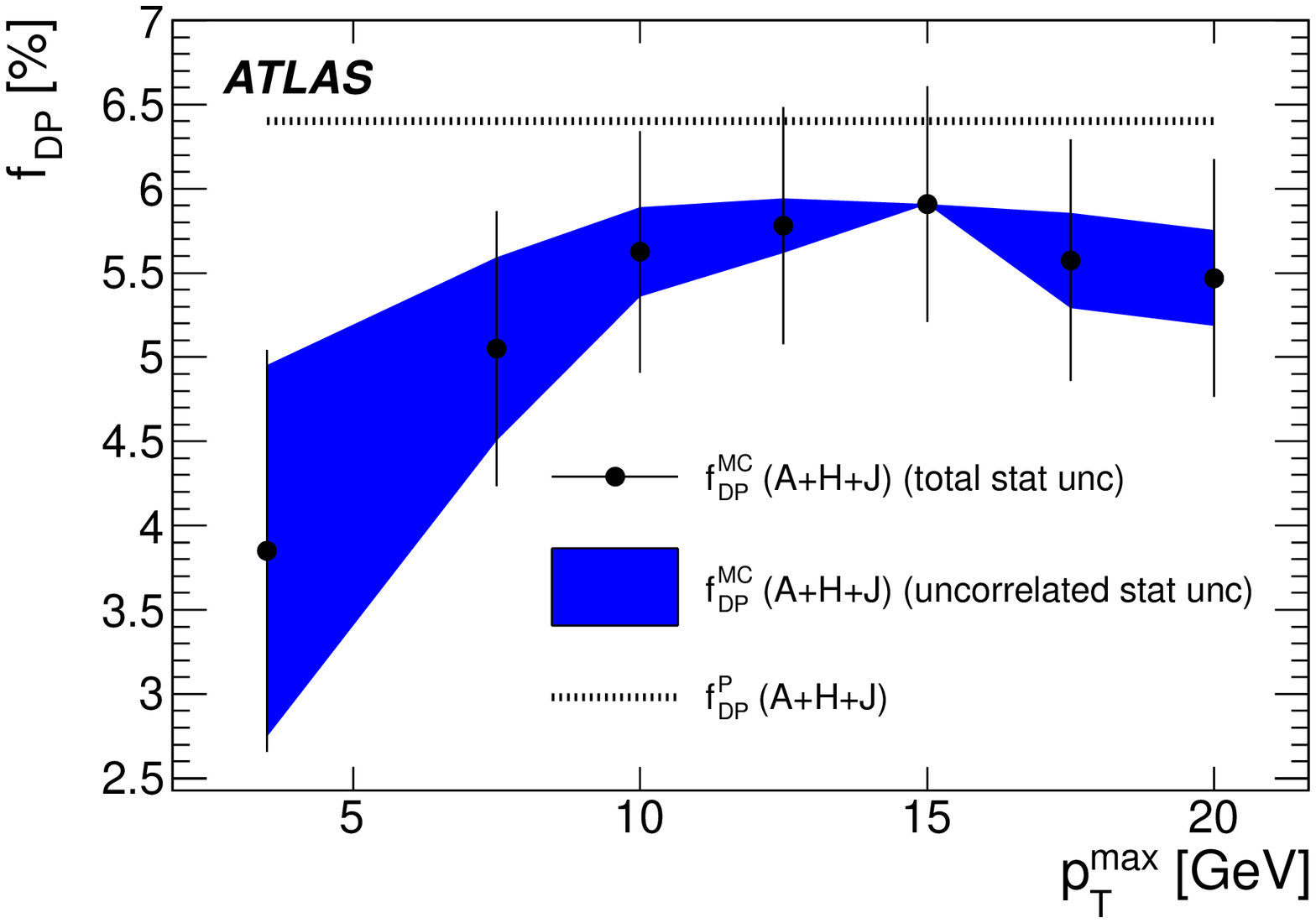}
\caption{Variation of the extracted fraction of double-parton scattering from the $\DeltaJn$ distribution in the \AHJ MC simulation, $\fDP{MC}$ (black points), as a function of the transverse momentum cut imposed on the scattered partons, $\ptmax$. The band illustrates the statistical component of the uncertainty of $\fDP{MC}$, relative to the reference sample with $\ptmax = 15\;$GeV, estimated by subtracting the statistical uncertainty of the reference sample in quadrature. The value, $\fDP{P}$, of the fraction of double-parton scattering obtained directly at the parton level (see \Eqref{fdp_partonlevel}) is also shown as a dashed line.}
\label{figure_partoncutoff}
\end{figure}

In this particular analysis, it is desirable to include these soft 
MPI partons in both templates.  In Template $B$ they are present by default,
and in Template $A$ they must also be allowed as they do form a contribution 
to the direct production of $W+2$-jet events in nature, which Template $A$ is set up
to model.   However, there is also an upper constraint on 
this threshold, since $W_{0\rm{j}}+2\rm{j_{DPI}}$ events where the jets have \pt~above 20\;GeV should not enter 
Template $A$. 
To determine the optimal $\ptmax$, alternative predictions 
for Template $A$ with different $\ptmax$ values were constructed from the inclusive
\AHJ sample.  Fits to $\DeltaJn$ distributions with different 
values of $\ptmax$ in the \AHJ sample forming Template $A$ were obtained.  At low 
$\ptmax$ values, $\leq$~7.5\;GeV, the fits do not show good agreement 
with the \AHJ distribution, since then partons from soft MPI are removed in Template $A$ and the hard jets become more correlated.  This 
leads to an underestimation of $\fDP{MC}$.  Conversely, at high $\ptmax$, 
genuine $W_0+2\rm{j_{DPI}}$ events are not removed when constructing Template $A$, 
also leading to an underestimation of the extracted $\fDP{MC}$. 
These two competing effects are visible in \Figref{figure_partoncutoff}, which 
shows $\fDP{MC}$ extracted as $\ptmax$ is varied, along with the value of 
$\fDP{P}$, as defined in \Eqref{fdp_partonlevel}.  
Therefore, the fit with the best agreement between $\fDP{P}$ and $\fDP{MC}$ 
determines the value of $\ptmax = 15\;$GeV chosen as default for the 
construction of Template~$A$ in this analysis.  
In this case, the value of $\fDP{P}$, as defined in \Eqref{fdp_partonlevel} (and indicated by the dotted line in \Figref{figure_partoncutoff}), is within 10\% of the value $\fDP{MC}(\DeltaJn) = 0.059\pm 0.007$ 
obtained from the templates fit. This implies that, when using the optimal $\ptmax$ derived above, the value of $\fDP{MC}$ (and therefore $\fDP{D}$) can be regarded as a measurement at the parton level with an associated 10\% uncertainty.

\subsection{Model dependence of Template $A$}
\label{SherpaComparison}
\begin{figure*}[t]
\centering
\mbox{  
\subfigure[]{\includegraphics[width=0.5\textwidth]{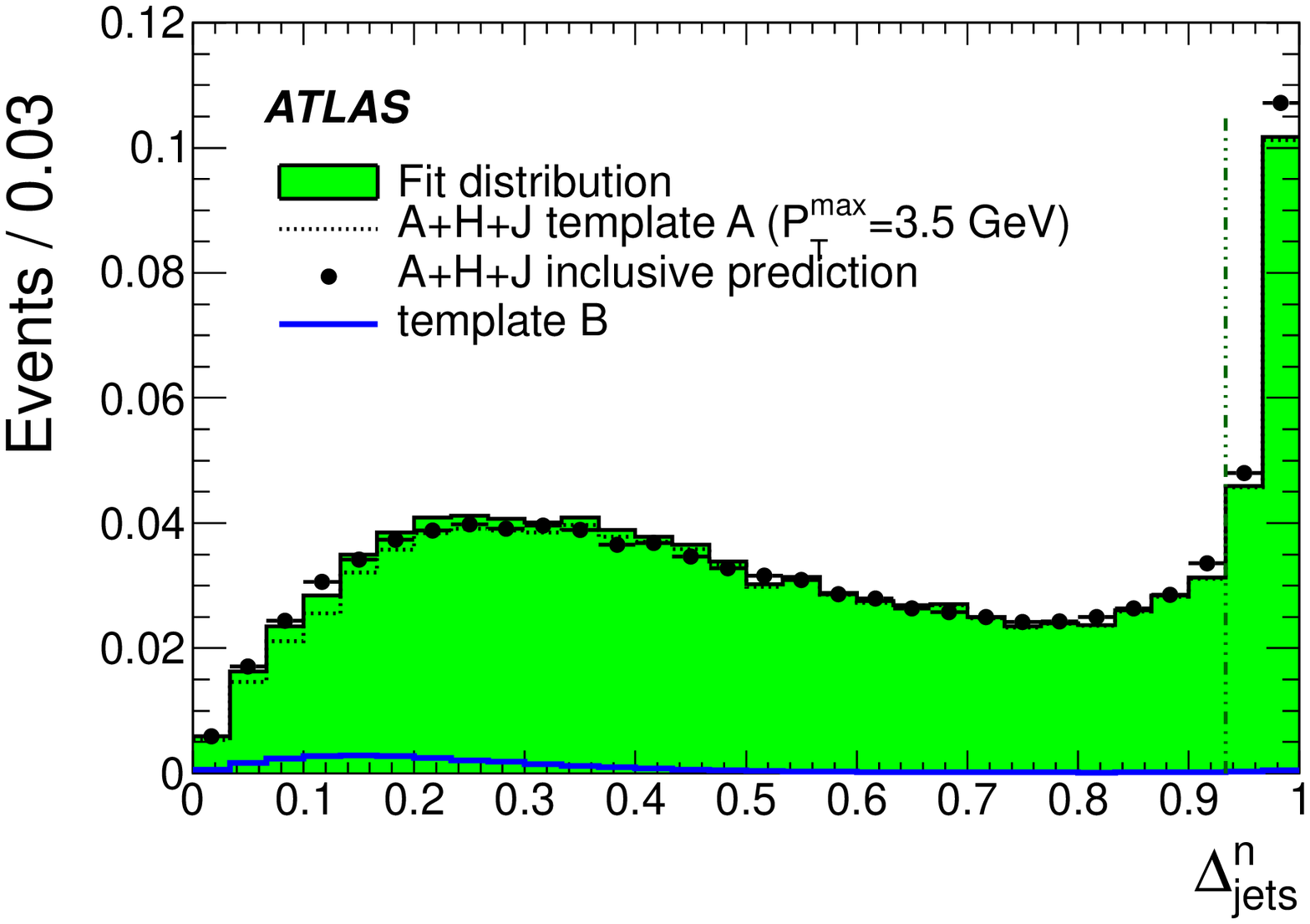}} \quad
\subfigure[]{\includegraphics[width=0.5\textwidth]{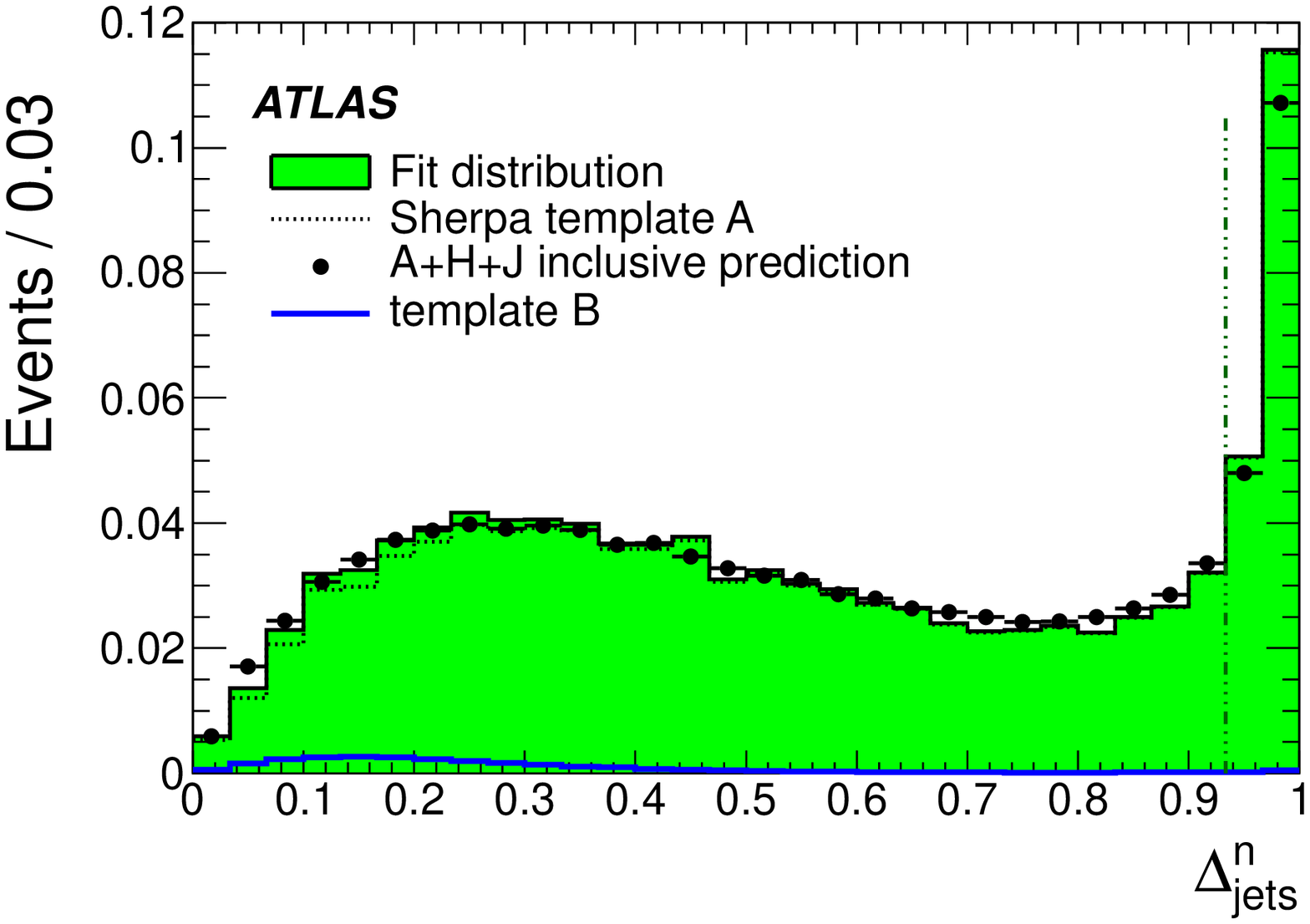}}
}
\caption{Distribution of $\DeltaJn$ in the inclusive \AHJ pseudo-data (dots). (a) \AHJ with $\ptmax = 3.5\;$GeV and (b) \textsc{Sherpa} were used to provide Template $A$. Other details are as in the caption to \Figref{figure_closurefit}.}
\label{figure_fit_deltaJets_3_5}
\end{figure*}

The effect of generator modelling is found by comparing the results for 
$\fDP{MC}$, when \textsc{Sherpa} is used to provide Template $A$, 
\fDP{S}, rather than \AHJ, \fDP{AHJ}. In this case for the \AHJ sample,
the $\ptmax$ parameter was set to $3.5\;$GeV to keep the samples comparable.  The results of these fits, 
after applying the \pileup correction $r_\mathrm{\pileup}$, are 
$\fDP{S} = 0.031\,\pm\,0.008$ and $\fDP{AHJ} = 0.034\,\pm\,0.006$. The quoted uncertainties are statistical and correlated between the two. However, the difference is smaller than either of the uncertainties and thus the results are consistent. The results of the two fits 
to the inclusive \AHJ sample are shown in \Figref{figure_fit_deltaJets_3_5}.

%% file: measurement.tex
\begin{figure*}[t]
\centering
\mbox{
\subfigure[]{\includegraphics[width=0.5\textwidth]{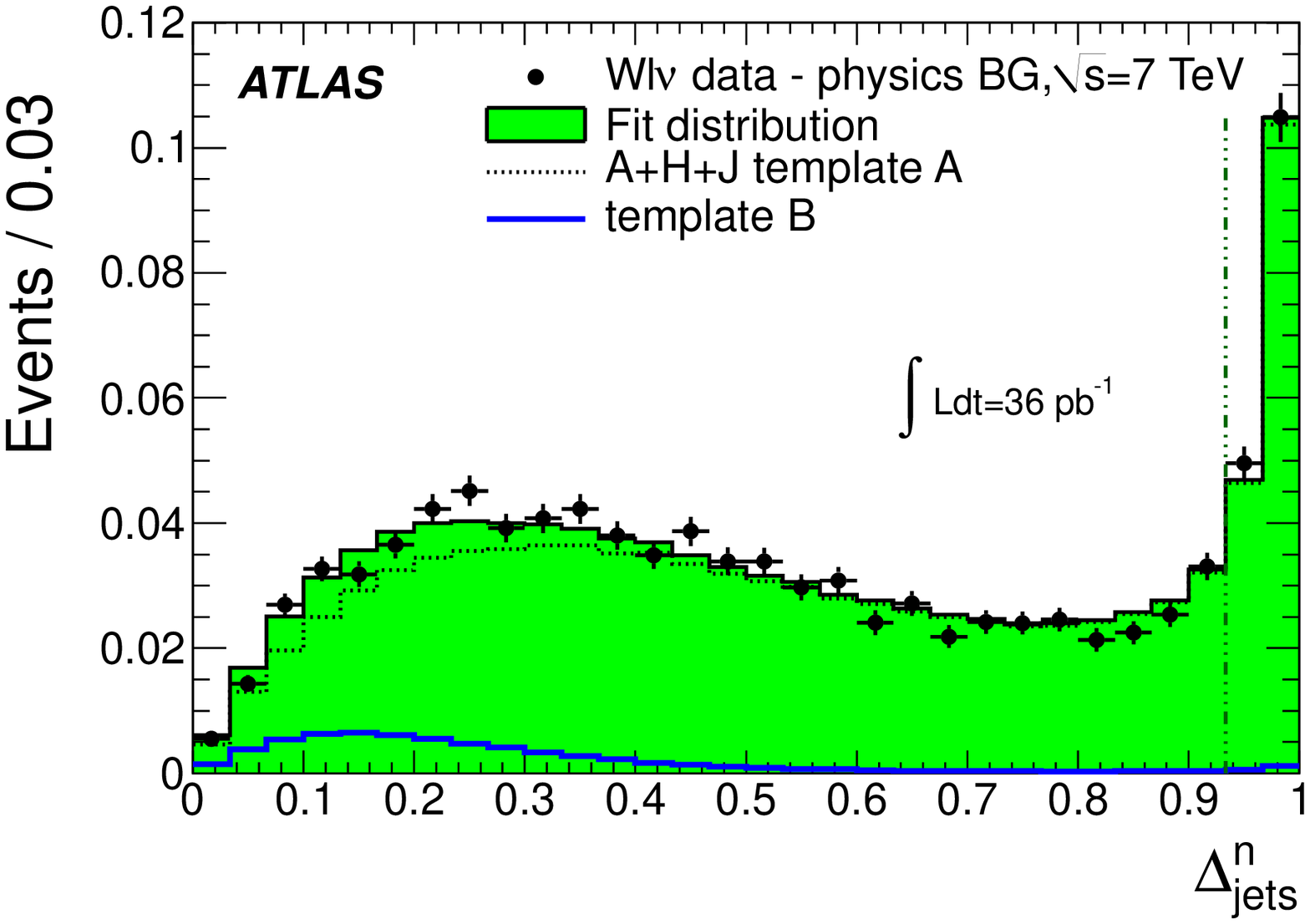}}
\subfigure[]{\includegraphics[width=0.5\textwidth]{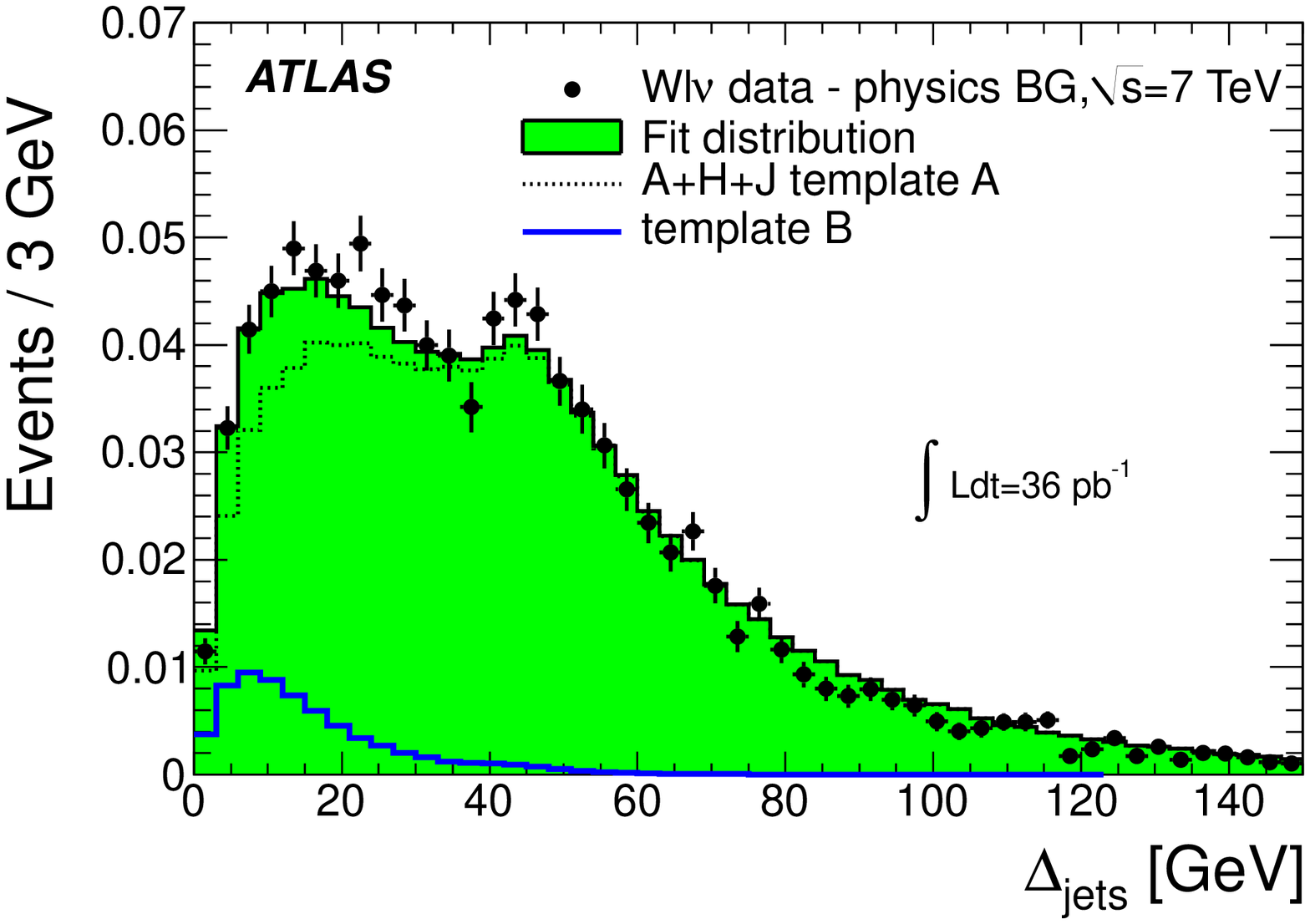}}
}
\caption{Distribution of  (a) $\DeltaJn$ and  (b) $\DeltaJ$ in the 
  background-subtracted data (dots) compared to the result from the best
  fit for $\fDP{D}(\DeltaJn)$. The result is shown as the green histogram.  In (a), the bins to the right of the vertical dash-dotted line were excluded from the fit.  Data and the overall fit have been normalised to unity, Template $A$ (dashed line) to $1-\fDP{D}(\DeltaJn)$ and Template 
  $B$ (blue solid line) to $\fDP{D}(\DeltaJn)$.}
\label{figure_fit}
\end{figure*}

Turning now to the data itself, the backgrounds are subtracted using the MC and data-driven estimates of \Secref{section_bg}. The parameter $\fDP{D}$ was extracted from a fit to the distribution of $\DeltaJn$ in the data sample of $W$+2-jet events after this physics background subtraction.  After applying the \pileup correction 
$r_\mathrm{\pileup}$, this yields

\begin{equation}
  \fDP{D}(\DeltaJn)\;=\;0.076\pm 0.013\;\;\mathrm{(stat.)},
\end{equation}

\noindent with $\chi^2/N_{\rm{dof}} = 37/28$. The result quoted
above is in good agreement with the result of a fit to the distribution of $\DeltaJ$.  The
resulting distributions obtained with $\fDP{D}(\DeltaJn)$ are shown in
\Figref{figure_fit}.

The systematic uncertainties on this extracted value of \protect$\fDP{D}$ are 
discussed in the following. 
\linebreak
\linebreak
\noindent {\bf Theoretical uncertainty}
 
\noindent The uncertainty due to generator modelling is estimated by comparing the results 
for $\fDP{S}$ with $\fDP{AHJ}$. To be conservative, the statistical uncertainty on $\fDP{S}$ is propagated as a symmetric systematic uncertainty on $\fDP{D}$ due to the modelling. An additional uncertainty is due to the choice of $\ptmax$. The systematic uncertainty due to the variation of this value is obtained by 
demanding that $\fDP{MC}$ and $\fDP{P}$ are consistent within statistical 
uncertainties.  From \Figref{figure_partoncutoff}, upward and downward variations of 
$\ptmax$ by 2.5\;GeV and 5\;GeV, respectively, are deduced.  The value of $\ptmax$ was varied by these two amounts, resulting in a variation in $\fDP{MC}$ of 0.003. 
The two uncertainties discussed above were added in quadrature to estimate the total theoretical uncertainty of 0.004.
\linebreak
\linebreak
\noindent {\bf Jet energy scale and resolution}

\noindent The overall impact of the jet energy scale on $\fDP{D}$ was determined by 
shifting the jet energy upwards and downwards in the MC samples by the jet energy 
scale uncertainty~\cite{jetpaper} and repeating the fit.  The variations were found to be 
+0.009 and $-$0.008, respectively. The larger of these two was symmetrised to provide the systematic uncertainty on $\fDP{D}$ due to that on the jet energy scale. Similarly, the overall impact of the jet energy resolution on 
$\fDP{D}$ was determined by degrading the jet energy resolution in the Monte 
Carlo samples by the jet energy resolution uncertainty, and re-performing the 
fit. The variation in $\fDP{D}$ in this case, assumed to be symmetric, was found to be $0.005$. 
\linebreak
\linebreak
\noindent {\bf Physics backgrounds and lepton response}

\noindent The impact of both physics background modelling and lepton response was 
considered via a direct comparison of $\fDP{D}$ obtained separately in the $\Wen$ and 
$\Wmn$ channels. Half of the obtained difference, which was at the sub-percent level, was taken 
as a measure of the associated uncertainty. 
\noindent The uncertainty associated with background subtraction was also determined by varying the background normalisation and shape.  The multi-jet background shape was varied in both 
channels by using MC instead of data-driven methods to estimate it, with no reversal of cuts. The 
shift in $\fDP{D}$ in this case was $-$0.001. 
The normalisation of the multi-jet background was varied in the $\Wen$ channel by taking 
the relative normalisation of the background contribution from two independent 
methods of background estimation -- this led to a relative 
variation of about 50\%.  For the $\Wmn$ channel, in which multi-jet background is better understood, a fractional shift of 20\% was assumed in the background normalisation.  The shift in 
$\fDP{D}$ when the multi-jet background normalisation was increased was found to be +0.008.  The uncertainty associated 
with the normalisation of electroweak and top backgrounds was evaluated by 
increasing the predicted \xsecs of these processes by their theoretical 
uncertainty of 5\%~\cite{WjetsXS}, resulting in a +0.001 variation in 
$\fDP{D}$. 
Symmetrising and then adding these uncertainties in quadrature yields the total 
uncertainty due to physics backgrounds and lepton response on $\fDP{D}$ of 0.008.
\linebreak
\linebreak
\noindent {\bf \Pileup}

\noindent The systematic uncertainty due to this effect was evaluated by 
propagating the statistical uncertainty on the \pileup correction, $r_\mathrm{\pileup}$, as 
determined in \Secref{sec:pileupcorr}, resulting in an \\uncertainty on $\fDP{D}$ of 0.010. \\
\linebreak
\noindent {\bf Impact of $\bm{W+1}$ jet}

\noindent It was verified that the contribution of $W+1$-jet configurations from the 
hardest scatter supplemented with a single jet from a secondary scatter formed a 
negligible contribution to the DPI rate at parton level, as well as in the 
modelling of Template $A$.
\linebreak

The individual contributions to the systematic uncertainty on $\fDP{D}$ are 
summarised in \Tabref{table_sys_fdp}. The contributions are added in quadrature, yielding 

\begin{equation}
\fDP{D} = 0.076 \pm 0.013~(\rm{stat.}) \pm 0.018~(\rm{sys.}). 
\end{equation}

\noindent The extracted value of $\fDP{D}$ is consistent with the value of $\fDP{MC}$ extracted in \Secref{sec:MCVal}, within the quoted uncertainties and hence with $\fDP{P}$ at the parton level, as discussed in \Secref{sec:partonlevel}. This implies that the MC models studied in the analysis describe the rate and kinematics of the DPI contribution well.

\begin{table}[htbp]
\centering
\caption{Summary of the fractional uncertainties on $\fDP{D}$. 
\label{table_sys_fdp}}
\begin{tabular}{ll}
\toprule
Systematic source & Uncertainty  [\%] \\ \hline
Theory & 10 \\ \midrule
\Pileup & 13 \\
Jet energy scale & 12 \\
Jet energy resolution & ~8 \\
Background modelling \& lepton response & 11 \\ \hline
Total systematic & 24 \\ 
Total statistical & 17 \\ 
\hline
\end{tabular}	
\end{table}

%% file: hadronlevel.tex
\begin{figure*}[htbp]
\centering
\mbox{
\subfigure[]{\includegraphics[width=0.5\textwidth]{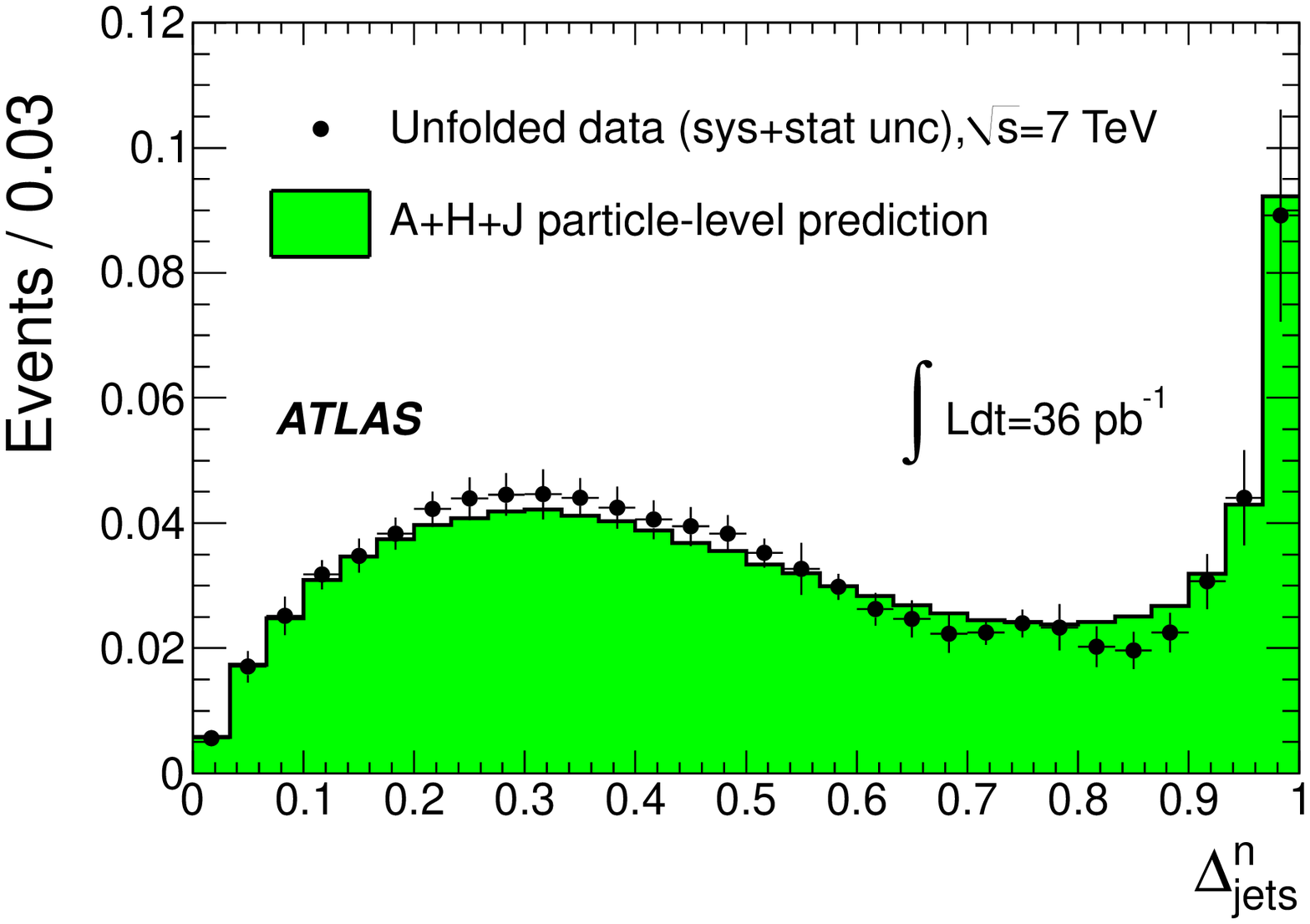}}\quad
\subfigure[]{\includegraphics[width=0.5\textwidth]{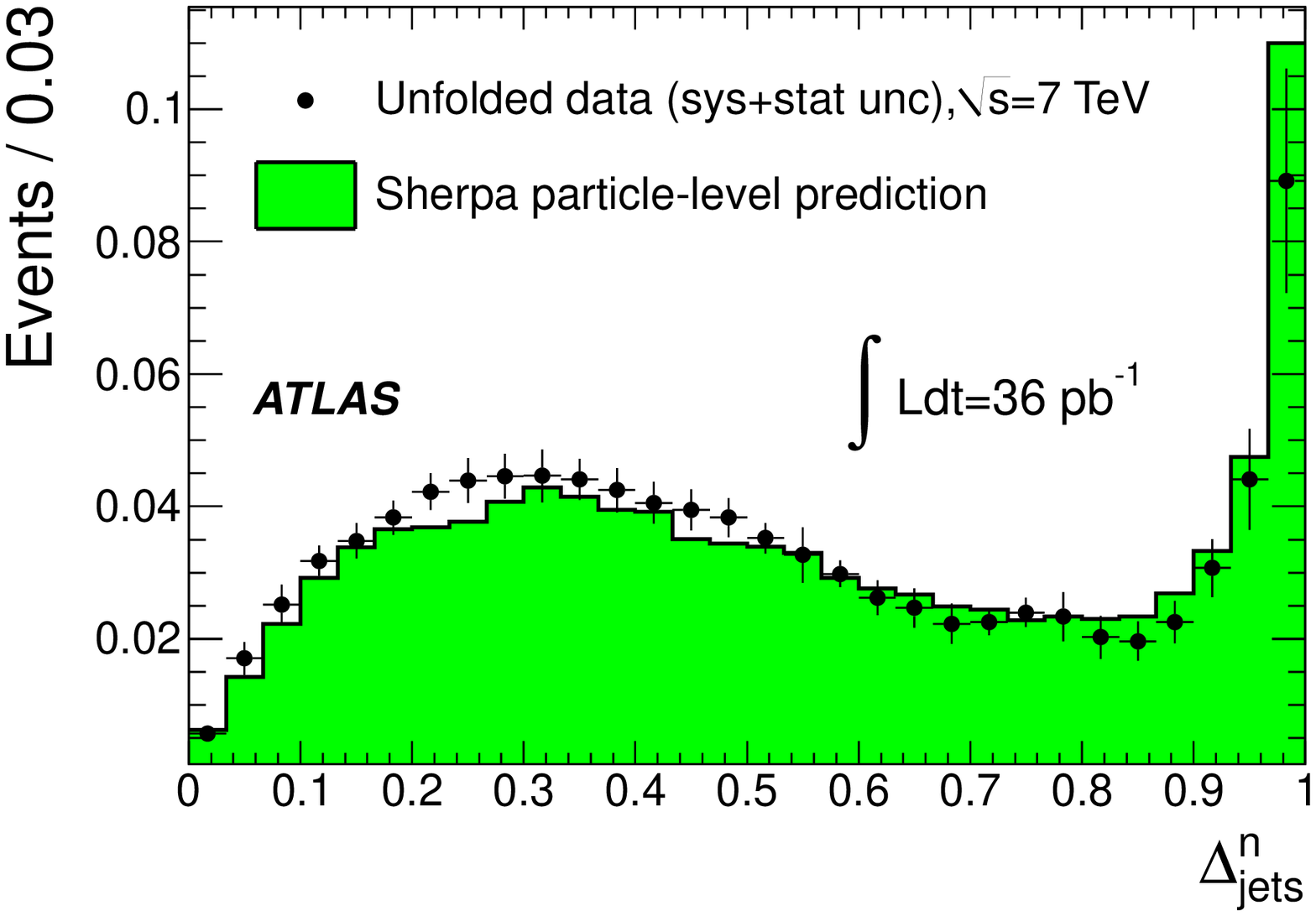}}
}
\mbox{
\subfigure[]{\includegraphics[width=0.5\textwidth]{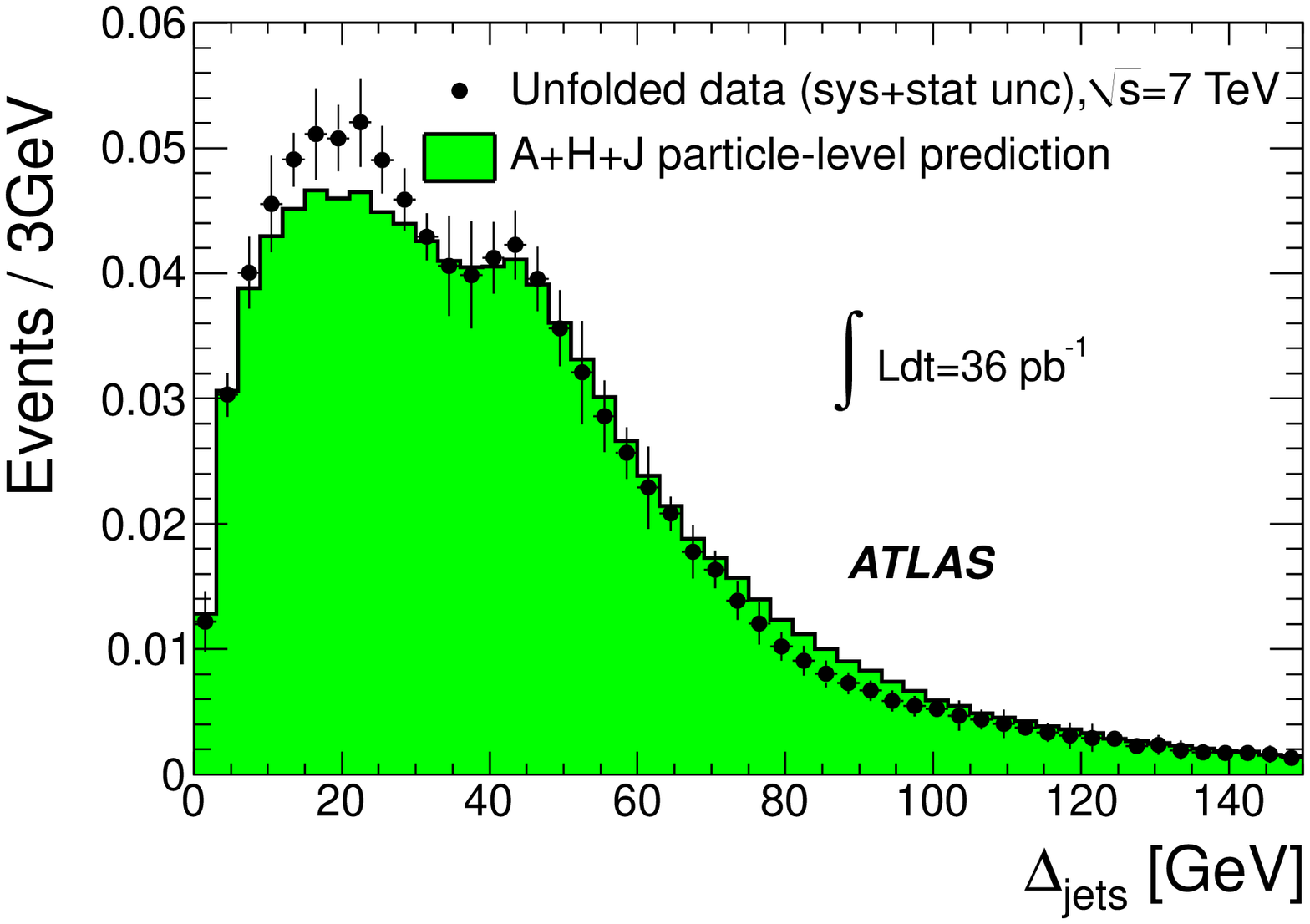}}\quad
\subfigure[]{\includegraphics[width=0.5\textwidth]{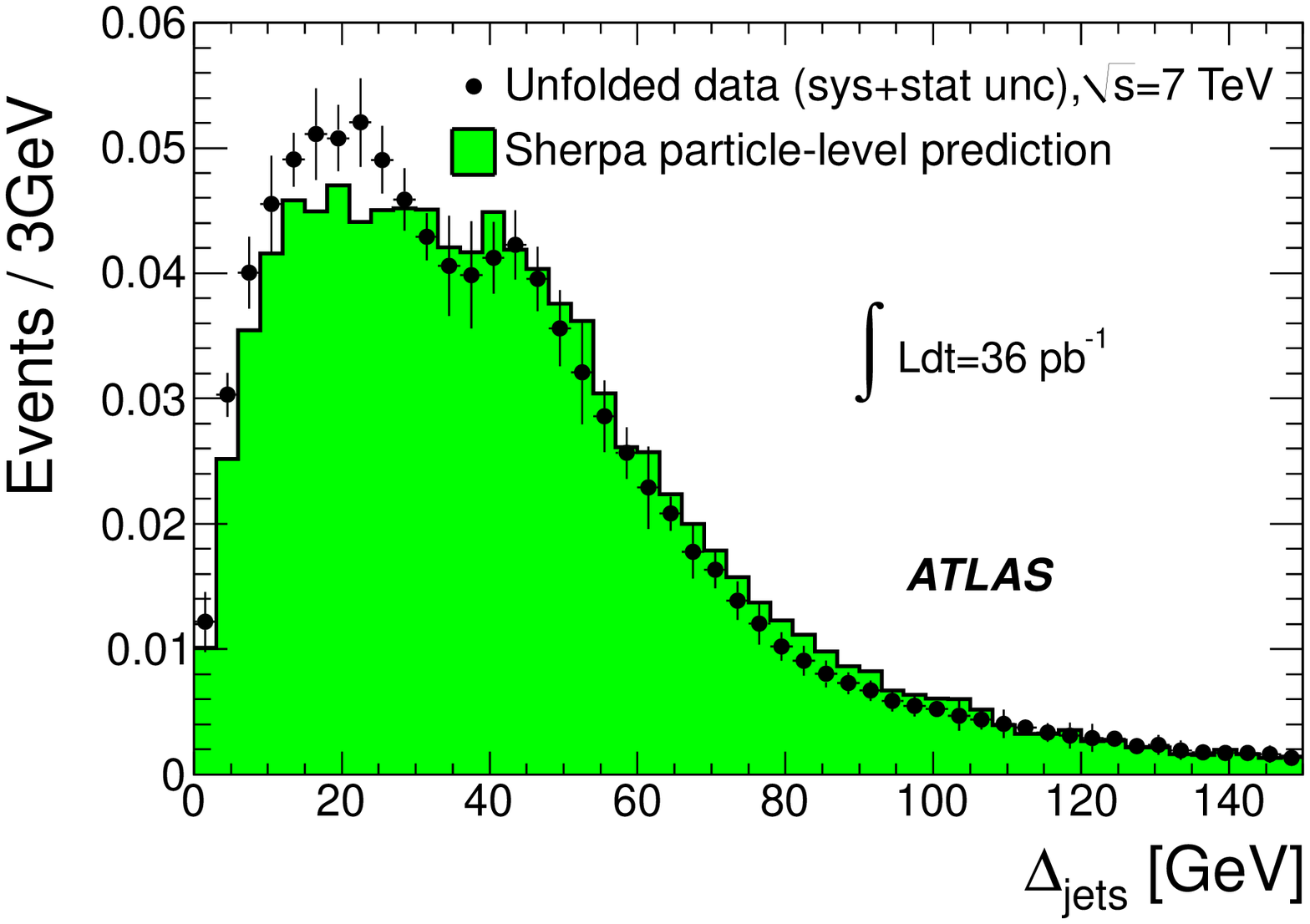}}
}
\caption{Distributions of (a, b) $\DeltaJn$ and 
  (c, d) $\DeltaJ$ in the data after unfolding to hadron level (dots) 
  compared to MC expectations from (a, c) \AHJ and (b, d) \textsc{Sherpa} at the hadron level (green histogram). The error bars represent the quadrature sum of systematic and statistical uncertainties on each bin, and both histograms have been normalised to unity.}
\label{figure_truthdists}
\end{figure*}

Since $\sigeff$ and $\fDP{P}$ are intrinsically parton-level quantities, they 
are not directly observable.  Conversely, $\fDP{D}$ contains residual 
dependencies on detector resolutions and efficiencies.  This renders a direct 
comparison with theoretical models impossible.  To allow the results of this
study to be used for comparisons with MPI models in the future, the key 
distributions have been corrected for detector effects to the final-state 
hadron level. 
The hadron-level requirements mirror the selection described in 
\Secref{sec:evsel}, except that cuts were applied to hadron-level quantities 
in MC simulation.  Hadron-level jets were constructed by running the same jet 
finder as for the detector level, using all final-state hadron-level 
particles with lifetimes longer than 30 ps as input, with the neutrinos and 
the charged leptons originating from the decay of the $W$ bosons being
excluded.  Jets were defined with the anti-$k_t$ algorithm with $\pt>20\;$GeV, 
$|y|<2.8$ and $R = 0.4$.  In addition, jets within a distance of
$\Delta R=0.5$ from the leptons were removed.  Dijet events were required to contain exactly 
two jets, reconstructed using the same algorithm, input objects and kinematic 
selection as already described.
The $\DeltaJn$ and $\DeltaJ$ distributions in $W+$2-jet data unfolded to 
the hadron level are shown in \Figref{figure_truthdists}.  The 
background-subtracted data has been corrected, using a Bayesian unfolding algorithm, 
to the hadron level using the RooUnfold package \cite{RooUnfold}.  The response 
matrix used to unfold the data was trained on \AHJ predictions and two 
iterations were used to converge to the unfolded distributions, resulting in a smoother distribution than that seen at detector level.
The unfolded results are compared with both \AHJ 
and \textsc{Sherpa} MC predictions directly obtained at hadron level. 

\begin{figure*}[htbp]
\centering
\mbox{
\subfigure[]{\includegraphics[width=0.5\textwidth]{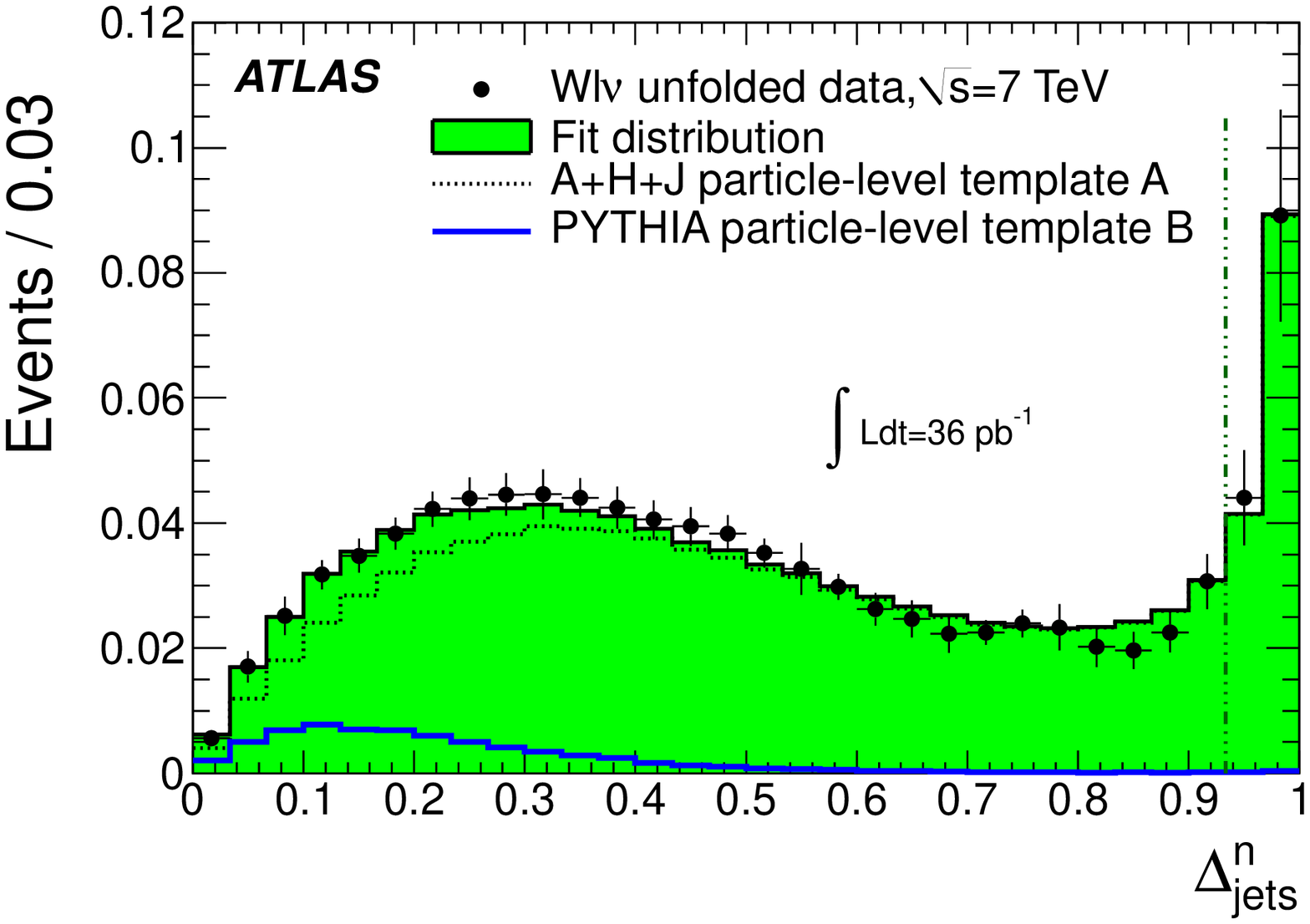}}\quad
\subfigure[]{\includegraphics[width=0.5\textwidth]{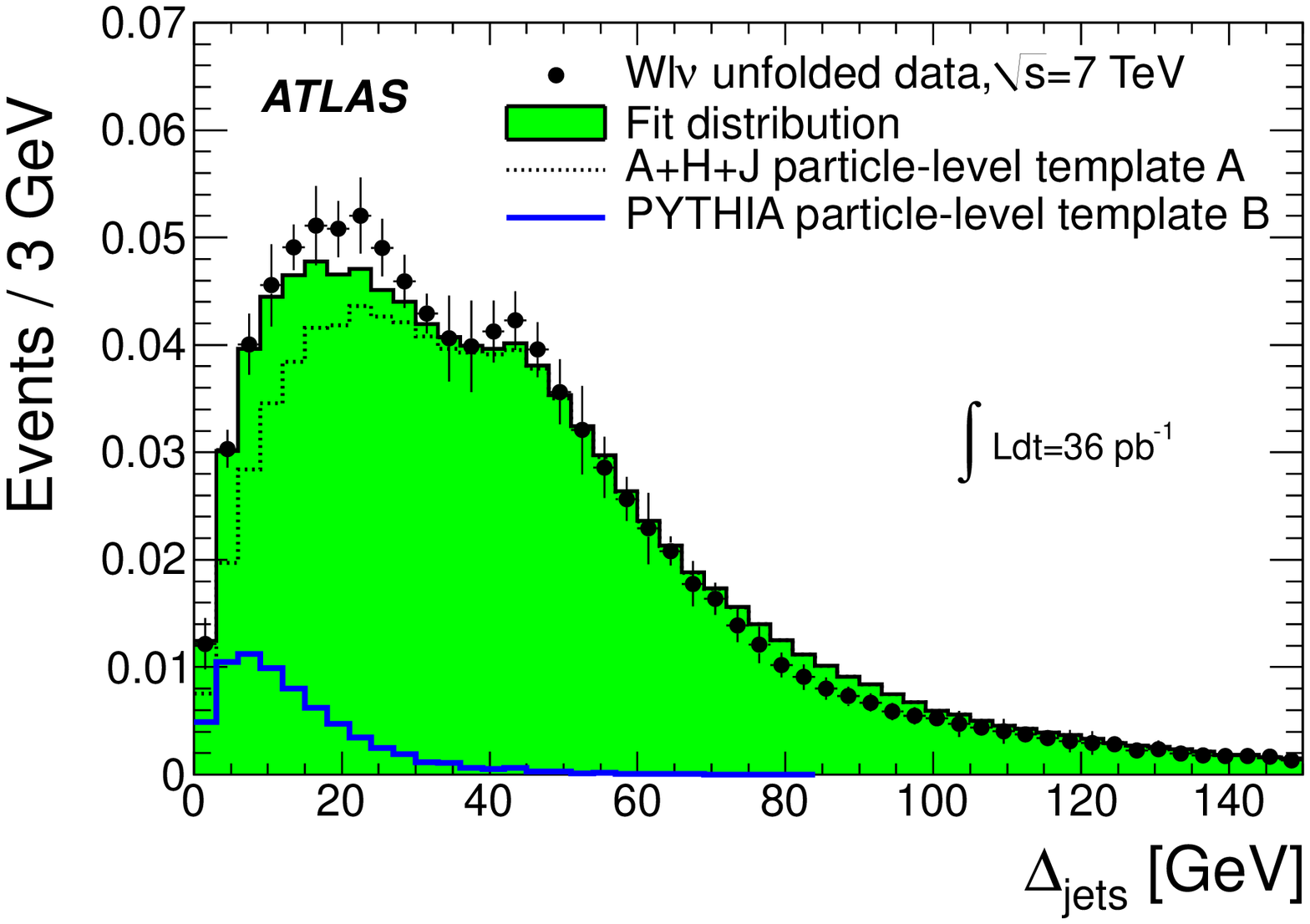}}\quad
}
\caption{Distributions of (a) $\DeltaJn$ and (b) $\DeltaJ$ in the data after unfolding to hadron level (dots) compared to the results of a linear combination with $\fDP{D}$ (green histogram) of Template $A$ 
  extracted from \AHJ hadron-level simulation (dashed line) and of Template $B$ obtained from 
  the \textsc{Pythia6} hadron-level simulation (solid blue line).  The error bars on the data 
  represent the quadrature sum of the statistical and systematic uncertainties. Data and the overall fit have been normalised to unity, Template $A$ to $1-\fDP{D}(\DeltaJn)$ and Template 
  $B$ to $\fDP{D}(\DeltaJn)$.}
\label{figure_truthfit}
\end{figure*}

The systematic uncertainties on the $\DeltaJn$ and $\DeltaJ$ distributions were obtained by repeating the studies outlined in \Secref{sec:measurement} with the exception of \pileup uncertainty. The latter was estimated by comparing the 
background-subtracted, corrected data distributions measured at hadron level 
with that obtained when rejecting all events other than those with exactly 
one primary vertex selected as described in \Secref{sec:evsel}.
The uncertainty due to the unfolding procedure itself was estimated 
as the shift in the corrected data distribution when \textsc{Sherpa} instead 
of \AHJ was used to train the response matrix.  
The overall uncertainty on the 
unfolded distribution was found, per bin of the distribution, by a quadrature 
sum of the uncertainties described above and is dominated by the \pileup 
uncertainty.
For completeness, the unfolded $\DeltaJn$ and 
$\DeltaJ$ distributions are compared with a linear combination of Template $A$ from \AHJ and Template $B$ from \textsc{Pythia6}, both at the hadron level, in proportions determined by the value of $\fDP{D}$, as shown in \Figref{figure_truthfit}. Perfect agreement between $\fDP{D}$ and its hadron-level equivalent is not expected as the phase space at hadron level, covered by the $W$+2-jet sample at detector level used for the determination of $\fDP{D}$, is not exactly the same as for the unfolded distribution. The value of $\fDP{D}$ 
determined directly via a fit at the hadron level was found to be within 10\% of the value determined at detector level.
%Clearly, the agreement of the unfolded data with this combination of the two templates
%is not as good as in \Figref{figure_fit}, at the detector level, due to
%acceptance effects that have not been properly accounted for here.  In that
%respect one could interpret the fairly good agreement of the unfolded data 
%and the templates in \Figref{figure_truthfit} as a sign for the anticipated
%smallness of these acceptance effects.  On the other hand, it is reassuring 
%to see that the dijet data can be safely replaced with a MC sample
%without hampering the overall agreement too badly.

%% file: sigeff.tex
The value of $\sigeff$ is related to $\fDP{D}$ through \Eqref{equation_sigeff}. The additional input of the exclusivity ratio, 
$N_{W_{0\mathrm{j}}}/N_{W+2\rm{j}} =23$, is evaluated from the event yields in the
selected $W+$2-jet and $W+$0-jet samples. The associated statistical uncertainty is at the 2\% level. Additionally, the number of dijet events $N_{2\rm{j}} = 9488$ is obtained from the event yield in 
the early period of 2010 data taking, corresponding to an integrated luminosity of $\mathscr{L}_{2\rm{j}}=184~\upmu \rm{b}^{-1}$. In this 
period of data taking, the trigger selection for dijet events was measured 
to be fully efficient ($\epsilon_{2\rm{j}}=1$)~\cite{jetpaper}.  
A further correction to $N_{2\rm{j}}$ was made for lepton--jet overlap removal, which was applied to the jets when constructing Template $A$ but not in Template $B$. It was evaluated by applying overlap removal for jets in Template $B$, giving a correction factor of 0.96.

The systematic uncertainties are summarised in \Tabref{table_sys_XS}, with the
following breakdown of their origins:
\begin{itemize}
\item The uncertainties on $\fDP{D}$ -- determined in \Secref{sec:measurement} -- are propagated asymmetrically to $\sigeff$.
\item Physics backgrounds and response to leptons -- the impact of the lepton energy response and background normalisation uncertainties were considered by propagating the uncertainty on $\fDP{D}$ obtained for this effect. In addition, the impact of lepton scale and the background normalisation uncertainties on the exclusivity ratio, $N_{W_{0\rm{j}}}/N_{W+2\rm{j}}$, were included.
\item Acceptance and response cancellation -- \Eqref{equation_effCancellation} is the result of factorisation which implies that the kinematics of the lepton and DPI system are not correlated, either in terms of geometrical acceptance or through detector response. Apart from the effect of lepton--jet overlap removal, which was discussed above, various sources of uncertainty on this assumption were considered and found to be negligible.
\end{itemize}

\begin{table}[htbp]
\centering
\caption{Summary of fractional systematic uncertainties on $\sigeff$. 
  \label{table_sys_XS}}
\begin{tabular}{ll}
\toprule
Systematic source & Uncertainty [\%] \\ 
\hline
$\fDP{D}$ & 24 \\
Background \& lepton response & 5  \\
Luminosity & 3 \\  
\hline
Total systematic & $^{+33}_{-20}$ \\ 
Total statistical & 17 \\ 
\hline
\end{tabular}	
\end{table}

The above leads to a measured central value of 
\begin{equation}
\sigeff(7~\,\mbox{\rm TeV}) = 15 \pm 3~(\rm{stat.})~^{+5}_{-3} ~(\rm{syst.})~mb.
\end{equation}

\begin{figure}[htbp]
\begin{center}              
\includegraphics[width=0.8\textwidth]{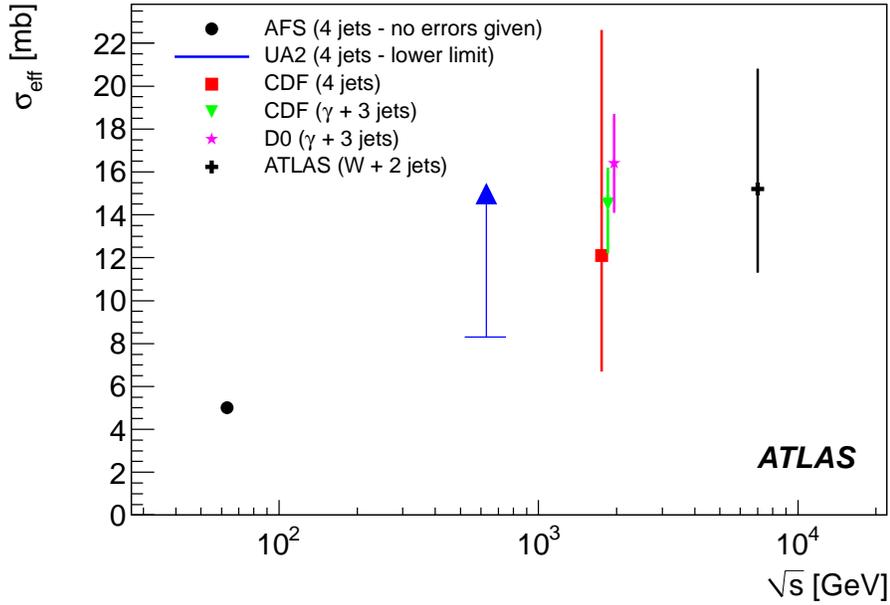}
\end{center}
\caption{The \com energy, $\sqrt{s}$, dependence of $\sigeff$ extracted 
             in different processes in different experiments. An offset has been applied to the 1.8~TeV data points in order to distinguish them.
             The error bars on the data points represent the statistical 
             and systematic uncertainties added in quadrature.}
\label{Fig:sigeff_with_s}
\end{figure}

Assuming factorisation, this value of $\sigeff$ is consistent with values previously measured in other experiments at lower centre-of-mass~\cite{Akesson:1986iv,Alitti:1991rd,Abe:1993rv,Abe:1997xk,Abazov:2009gc}, as can be seen in \Figref{Fig:sigeff_with_s}.

%% file: conclusions.tex
The \dblp interaction rate $\fDP{D}$ in events with a $W$ boson and exactly two jets in the final state has been extracted in $pp$ collisions at a \com energy of $\sqrt{s} = 7~\TeV$ using data corresponding to an integrated luminosity of 36~\ipb. For jets with transverse momentum $p_{\rm T} > 20$~GeV and rapidity $|y|<2.8$, in the ATLAS detector at the LHC, a central value of
\vspace{0.3cm}\\
\centerline{$\fDP{D} = 0.08 \pm 0.01~(\rm{stat.}) \pm 0.02 ~(\rm{sys.})$}\vspace{0.3cm}
\noindent is obtained. In terms of measured rate and kinematics of the dijet system there is good agreement with the predictions of the MC models studied in the analysis.
The result for $\fDP{D}$ is used to extract the parameter $\sigeff$ through the production of $W+$2-jet
events. The value extracted from data is 
\vspace{0.3cm}\\
\centerline{$\sigeff(7~\,\mbox{\rm TeV}) = 15 \pm 3~(\rm{stat.})~^{+5}_{-3} ~(\rm{sys.})~mb.$}\vspace{0.3cm}
\noindent This value is consistent with values previously measured in other experiments at lower \com energies.

%% file: Acknowledgements.tex
\section{Acknowledgements}

% Standard acknowledgements start here
%----------------------------------------------
We thank CERN for the very successful operation of the LHC, as well as the
support staff from our institutions without whom ATLAS could not be
operated efficiently.

We acknowledge the support of ANPCyT, Argentina; YerPhI, Armenia; ARC,
Australia; BMWF and FWF, Austria; ANAS, Azerbaijan; SSTC, Belarus; CNPq and FAPESP,
Brazil; NSERC, NRC and CFI, Canada; CERN; CONICYT, Chile; CAS, MOST and NSFC,
China; COLCIENCIAS, Colombia; MSMT CR, MPO CR and VSC CR, Czech Republic;
DNRF, DNSRC and Lundbeck Foundation, Denmark; EPLANET, ERC and NSRF, European Union;
IN2P3-CNRS, CEA-DSM/IRFU, France; GNSF, Georgia; BMBF, DFG, HGF, MPG and AvH
Foundation, Germany; GSRT and NSRF, Greece; ISF, MINERVA, GIF, DIP and Benoziyo Center,
Israel; INFN, Italy; MEXT and JSPS, Japan; CNRST, Morocco; FOM and NWO,
Netherlands; BRF and RCN, Norway; MNiSW, Poland; GRICES and FCT, Portugal; MERYS
(MECTS), Romania; MES of Russia and ROSATOM, Russian Federation; JINR; MSTD,
Serbia; MSSR, Slovakia; ARRS and MVZT, Slovenia; DST/NRF, South Africa;
MICINN, Spain; SRC and Wallenberg Foundation, Sweden; SER, SNSF and Cantons of
Bern and Geneva, Switzerland; NSC, Taiwan; TAEK, Turkey; STFC, the Royal
Society and Leverhulme Trust, United Kingdom; DOE and NSF, United States of
America.

The crucial computing support from all WLCG partners is acknowledged
gratefully, in particular from CERN and the ATLAS Tier-1 facilities at
TRIUMF (Canada), NDGF (Denmark, Norway, Sweden), CC-IN2P3 (France),
KIT/GridKA (Germany), INFN-CNAF (Italy), NL-T1 (Netherlands), PIC (Spain),
ASGC (Taiwan), RAL (UK) and BNL (USA) and in the Tier-2 facilities
worldwide.

%% file: atlas_authlist.tex
% ATLAS Collaboration author list
% Data extracted on 10-Feb-2013 for paper reference STDM-2011-29
%\documentclass[11pt]{article}
%\usepackage{a4wide}\begin{document}
\begin{flushleft}
{\Large The ATLAS Collaboration}

\bigskip

G.~Aad$^{\rm 48}$,
T.~Abajyan$^{\rm 21}$,
B.~Abbott$^{\rm 111}$,
J.~Abdallah$^{\rm 12}$,
S.~Abdel~Khalek$^{\rm 115}$,
A.A.~Abdelalim$^{\rm 49}$,
O.~Abdinov$^{\rm 11}$,
R.~Aben$^{\rm 105}$,
B.~Abi$^{\rm 112}$,
M.~Abolins$^{\rm 88}$
% ...
\bigskip
\\
% ...
$^{11}$ Institute of Physics, Azerbaijan Academy of Sciences, Baku, Azerbaijan\\
$^{12}$ Institut de F{\'\i}sica d'Altes Energies and Departament de F{\'\i}sica de la Universitat Aut{\`o}noma de Barcelona and ICREA, Barcelona, Spain\\
% ...
$^{21}$ Physikalisches Institut, University of Bonn, Bonn, Germany\\
% ...
$^{48}$ Fakult{\"a}t f{\"u}r Mathematik und Physik, Albert-Ludwigs-Universit{\"a}t, Freiburg, Germany\\
$^{49}$ Section de Physique, Universit{\'e} de Gen{\`e}ve, Geneva, Switzerland\\
% ...
$^{88}$ Department of Physics and Astronomy, Michigan State University, East Lansing MI, United States of America\\
% ...
$^{105}$ Nikhef National Institute for Subatomic Physics and University of Amsterdam, Amsterdam, Netherlands\\
% ...
$^{111}$ Homer L. Dodge Department of Physics and Astronomy, University of Oklahoma, Norman OK, United States of America\\
$^{112}$ Department of Physics, Oklahoma State University, Stillwater OK, United States of America\\
% ...
$^{115}$ LAL, Universit{\'e} Paris-Sud and CNRS/IN2P3, Orsay, France\\
% ...
\end{flushleft}

%\end{document}
% Created with ./xml2latex.py